%% file: Durka-phd-thesis.tex
\documentclass[phd, 11pt, a4paper, inlinechaptertoc]{psuthesis}
\usepackage[left=2.5cm,top=4cm,right=2.5cm,bottom=3cm]{geometry} 
\usepackage{fancyhdr}
\usepackage{amsmath}
\usepackage{amssymb}
\usepackage{amsthm}
\usepackage{amsmath}
\usepackage{exscale}
\usepackage{dsfont}
\usepackage{fancybox}
\usepackage[mathscr]{eucal}
\usepackage{bm}
\usepackage{eqlist} 
\usepackage[final]{graphicx}
\usepackage[dvipsnames]{color}
\usepackage{bbding,hyperref}
\definecolor{maroon}         {cmyk}{0   , 0.93, 0.9   , 0.40}

\DeclareGraphicsExtensions{.pdf, .jpg}

\usepackage[Lenny]{fncychap}
\ChTitleVar{\Huge\sffamily\scshape}

\def\bbone{{\mathchoice {\rm 1\mskip-4mu l} {\rm 1\mskip-4mu l}
{\rm 1\mskip-4.5mu l} {\rm 1\mskip-5mu l}}}

\title{Deformed BF theory as a theory of\\
 gravity and supergravity
}

\author{Remigiusz Durka}
\dept{Institute for Theoretical Physics}
\degreedate{March 2012}
\copyrightyear{2012}

\documenttype{Thesis}

\submittedto{Institute for Theoretical Physics}

\numberofreaders{0}

\secondthesissupervisor{Second T. Supervisor}

\honorsdepthead{Department Q. Head}

\advisor[~~]
        {\textbf{Jerzy Kowalski-Glikman}}
        {Professor of Institute for Theoretical Physics, University of Wroclaw}

\readerone[Thesis Referee]
          {Krzysztof Meissner}
          {Professor of Institute for Theoretical Physics, University of Warsaw}

\readertwo[Thesis Referee]
          {Bogus\l aw Broda}
        {Professor of Institute for Theoretical Physics, University of Lodz}

\begin{document}
\frontmatter

%



\psutitlepage
\clearpage
\thispagestyle{empty}
\mbox{}
\clearpage




%
\pagestyle{plain}
\chapter{Abstract}
    \begin{singlespace}
        \input{Abstract}
    \end{singlespace}
\newpage


%
\chapter{Publications}
\begin{singlespace}
    \input{Publications}

\end{singlespace}

\chapter{Acknowledgments}
\begin{singlespace}
    \input{Acknowledgments}

\end{singlespace}



\thesistableofcontents

\thesismainmatter

\allowdisplaybreaks{
%

\part{Gravity and BF theory}

\pagestyle{fancy}
\fancyhead{}  
\fancyfoot{}  
\fancyhead[LE,RO]{\thepage}
\chapter{Introduction}
\section{Outline}

Although soon we will celebrate 100 years of General Relativity, we are still far from understanding the true nature of gravity, and finding the way leading to satisfactory unification of gravity with other interactions. Mainstream theories (string theory, loop quantum gravity) offer different approaches to describe gravity on a microscopic level and arrive at the quantum theory, but none of them is spectacularly successful, and all are troubled by many issues.

The problem might lie with the chosen variables or the assumptions made along the way, keeping us from seeing a bigger picture and the full resolution of our problems. We might be also missing something essential, like for example the necessity of a cosmological constant, or just take for granted some simplifications, which could neglect important pieces and conditions needed to complete our models.

Through decades physicists have tried to generalize, and extend the theory of gravity in numerous different ways. Some of the motivations originated in direct comparing gravity with other interactions, supersymmetric extensions, when other were simply subject of interest just from the formal side. We make step toward such considerations with the alternative model of gravity, and analysis of its features, offering wider framework than Einstein's General Relativity (GR). Proposed deformation of the $BF$ theory \cite{Freidel:2005ak}, on which we will be focusing in this thesis, contains very interesting structure behind it. One can use it to address many questions from a contemporary gravity research. 

To this end we will start with a generalization of the Einstein theory to the Cartan theory, and so called the first order (or tetrad) formulation \footnote{Notice, that in a literature one can often find incorrect name of the \textit{Palatini formalism} or the \textit{Palatini action}. Palatini proposal concerned the variation principle with the independent variations due to $g_{\mu\nu}$ and the metric connection $\Gamma^{\lambda}_{~\mu\nu}(g)$. Ellie Cartan was the first, who provided truly independent description with the pair of metric $g_{\mu\nu}$, and arbitrary (no longer 'metric') connection $\Gamma^{\lambda}_{~\mu\nu}$.}. It will admit, crucial for Cartan's philosophy, the independence of the metric and notion of the connection, but through solving vacuum field equations it reaches direct equivalence with standard GR. 

In the next chapter we will concentrate on the intriguing reformulation, being a key to all presented here results. MacDowell and Mansouri proposal \cite{MacDowell:1977jt} implements everything above into theory reproducing GR as an effect of the gauge symmetry breaking, and taking the form similar to the Yang-Mills theory. Its unique features will be explored in the context of the $BF$ model, characterized by even more intriguing appearance, and at the same time encoding the most general form of action for the first order gravity.

Second part will be devoted to the applications. It corresponds to the author's investigations in the several different contexts, like supergravity \cite{Durka:2009pf}, derivation of the gravitational Noether charges related with the Immirzi parameter (see \cite{Durka:2011yv} and \cite{Durka:2011zf}), the AdS--Maxwell group of symmetries \cite{Durka:2011nf, Durka:2011gm}, and the canonical analysis \cite{Durka:2010zx}. 

We will show, that the Immirzi parameter (often called Barbero-Immirzi) \cite{Barbero:1994ap, Immirzi:1996di, Holst:1995pc}, being one of the essential elements underlying in a foundations of loop quantum gravity, does not influence supergravity coming from the super-BF model, and in the Noether charge approach it does not alter the black hole thermodynamics for the standard cases of AdS-Schwarzschild and AdS-Kerr spacetimes, with intriguing impact only for the case of Taub-NUT spacetime. 

$BF$ theory can be also seen as a convenient platform to construct gravity and supergravity using modified Maxwell algebra, being interesting algebra extension.

\chapter{From Cartan theory to MacDowell-Mansouri gravity}

\section{Cartan theory}

The heart of General Relativity is the metric tensor $g_{\mu\nu}$, which describes the infinitesimal distance between the two nearby spacetime points $x^\mu$ and $x^\mu+dx^\mu$
\begin{equation}\label{ch2-1}
    ds^2=g_{\mu\nu} \,dx^\mu\,dx^\nu\,.
\end{equation} 
Second important ingredient is the parallel transport. It is described by the \textit{connection}, which says how much vector is varied along this transport, becoming crucial part of the covariant derivative replacing the notion of a partial derivative
\begin{equation}\label{ch2-2}
    \delta v^\alpha=-\Gamma^\alpha{}_{\mu\nu}\, v^\mu\, dx^\nu\,,\qquad\qquad\nabla_\mu
v_\nu=\partial_\mu
v_\nu-\Gamma^\lambda_{\;\mu\nu}v_\lambda\,.
\end{equation}
By the Einstein postulate, saying that antisymmetric part of the connection $ T^\lambda_{\mu\nu}=\Gamma^\lambda_{\mu\nu}-\Gamma^\lambda_{\nu\mu}$, called the torsion, is equal zero, the connection becomes a function of the metric $\Gamma^\lambda_{\mu\nu}(g_{\alpha\beta})$. With $\nabla_\lambda g_{\mu\nu}=0$ this uniquely determines symmetric part to be given by the Christoffel symbol
\begin{equation}\label{ch2-3}
\Gamma^\alpha_{\{\mu\nu\}}=\frac{1}{2}g^{\alpha\lambda} (\partial_\mu g_{\lambda\nu}+\partial_\nu g_{\mu\lambda}-\partial_\lambda g_{\mu\nu}) \,,\qquad\Gamma^\alpha_{[\mu\nu]}=0\,.
\end{equation}
Then, one uses such a connection $\Gamma^\rho{}_{\mu\nu}(g)$ to build the Riemann tensor
\begin{equation}\label{ch2-4}
    {R^\rho}_{\sigma\mu\nu}(g) = \partial_\mu\Gamma^\rho_{\nu\sigma}(g) - \partial_\nu\Gamma^\rho_{\mu\sigma}(g) + \Gamma^\rho_{\mu\lambda}(g) \Gamma^\lambda_{\nu\sigma}(g)  - \Gamma^\rho_{\nu\lambda}(g) \Gamma^\lambda_{\mu\sigma}(g) \,,
\end{equation}
the Ricci tensor $R_{\mu\nu}= {R^\rho}_{\mu\rho\nu}$, and the curvature scalar $R=g^{\mu\nu} R_{\mu\nu}=R^\mu{}_\mu$
employed to construct the Hilbert-Einstein action
\begin{equation}\label{Hilbert-Einstein}
\displaystyle S =  \frac{1}{16\pi G} \int \,\mathrm{d}^4 x\sqrt{-g} \,\left( R(g) - 2 \Lambda \right)+S_{matter}\,.
\end{equation}
The Einstein field equations derived from it,
\begin{equation}\label{ch2-5}
 R_{\mu\nu}-\frac{1}{2}g_{\mu\nu}R+\Lambda g_{\mu\nu}=8\pi G\, T_{\mu\nu}\,,
\end{equation}
provide profound and intimate relation between the curvature, and the energy density distribution in the spacetime. We have tested this theory to a large extend, in the end always confirming its predictions. However we still can't find a way to unite it with the other forces, and apply it to the scales beyond its applicability. To overcome this impasse we can try to extend the geometrical principles of General Relativity to microphysics allowing for desired comparison of these two worlds. 

To include microscale one unavoidably must take into account, that matter is not only characterized by the mass, but also by the spin distribution. Usual thinking about matter relates the mass/energy distribution with the energy-momentum tensor $T_{\mu\nu}$. On the other hand, the spin density tensor describes the spin distribution. One can think, that this object could be somehow related to some geometric quantity, similar to a way how the energy-momentum tensor is related to the curvature \cite{deSabbata:1994wi}. In fact, as we will see, it is related to the torsion, a quantity by far (due to the Einstein postulate) set always to be zero. In most of the physical objects spins are chaotic, so they are averaged to zero. That's why $T_{\mu\nu}$ is sufficient to describe dynamics in the vast majority of applications, and it is believed that the symmetric part of the connection is all we need.

Such a setting agrees with the argumentation of Ellie Cartan given even before the discovery of a spin. He argued with Einstein that the Riemannian manifold is in principle equipped with two independent quantities: the metric, and the connection being a notion of the parallel transport independent from the metric. This extends gravity to interesting class of torsion theories \cite{Hehl:1976kj}, \cite{Shapiro:2001rz}. They result in the one more kind of field equations, except the equations coming from the metric variation of the action with a curvature scalar replaced by an expression built from the both symmetric and antisymmetric connections, and using the antisymmetric energy-momentum tensor. With torsion $ T^\alpha_{\mu\nu}=\Gamma^\alpha_{\mu\nu}- \Gamma^\alpha_{\nu\mu}$ and spin density $ S^\alpha_{\mu\nu}$ tensors the variation of such a action over the torsion brings an equation first given by Sciama and Kibble
\begin{equation}\label{ch2-6}
    T^\alpha_{\mu\nu} - g^\alpha{}_{\mu}T^\rho_{\nu\rho} -g^\alpha{}_{\nu}T^\rho_{\rho\mu} = 8\pi G S^\alpha_{\mu\nu}\,.
\end{equation}

Torsion offering platform to incorporate spin into the structure of the General Relativity is very intriguing. However, in this thesis we will be restricted to vacuum gravity, so without fermionic matter, of course except the case of gravitino: the superpartner of a gravition in supergravity. That's why we will only investigate limited aspects, directly serving formal combining the torsion and the curvature together into one structure. Without the fermionic content the extended framework always reduces to General Relativity based on the simplest realization of the connection given by the Christoffel symbol.

\section{First order gravity}

Let us adopt Cartan point of view of the economy of assumptions, however not the economy of variables, and introduce formulation of gravity not represented by the metric, but with making a transition to the tangent space by the tetrad and the spin connection \cite{Trautman:2006fp}. This is done because this is the formalism in which supergravity has to be formulated \cite{Zanelli:2002qm}. 

Mapping between the spacetime manifold $\mathcal{M}$ and flat Minkowski tangent space $T_x$ is assured by the means of object called the tetrad (in 4D it's often named as \textit{vierbain}), which serves to represent tensors from the spacetime manifold by the tensors on the tangent space. The infinitesimal $dx^\mu$ on the $\mathcal{M}$ is mapped to corresponding $dx^a$ on $T_x$
\begin{equation}
    dx^a=e^a_\mu\, dx^\mu\,.
\end{equation}
allowing us to treat $e^a_\mu(x)$ as a local orthonormal frame on $\mathcal{M}$. Other example concerning the metrics in both spaces: 
\begin{equation}\label{ch2-7}
 \eta_{ab}=g_{\mu\nu}(x)\, e_a^\mu(x)\, e_b^{\nu}(x)\qquad \mbox{and}\qquad   g_{\mu\nu}(x)=\eta_{ab}\,e^a_\mu(x)\, e^b_{\nu}(x)\,,
\end{equation}
shows that one can easily find the metric from a given tetrad, and vice versa.
 
To make this work the tetrad should transform as a covariant vector under diffeomorphisms on $\mathcal{M}$ and as a contravariant vector under local Lorentz $SO(1,3)$ rotations of $T_x$: 
   $ e^a_\mu(x)\to e'{}^{a}_\mu(x)=\Lambda^a{}_b(x)e^b_\mu(x)$. Metrics should not change by these transformations, that's why we have condition $\Lambda^a{}_c(x) \Lambda^b{}_d (x)\, \eta_{ab} =\eta_{cd}$ with matrices $\Lambda(x)$ naturally forming $SO(1,3)$ group. Because the Lorentz group acts at each point separately we need to introduce a gauge field $\omega$ to compensate for comparing the tangent spaces for two different points. Therefore a covariant derivative $ D^\omega_\mu$ for tensors in $T_x$ is composed by the \textit{spin connection}\footnote{Name comes from the fact it allows to incorporate spinors, but better choice would be a Lorentz connection.} $\omega^{ab}_\mu$, in principle independent from the tetrad. As an example we give
\begin{equation}
    D^\omega_\mu X^{a}=\partial_\mu  X^{a}+\omega^a_{\mu\,b} \, X^b\,,\qquad    D^\omega_\mu Y^{ab}=\partial_\mu  Y^{ab}+\omega^a_{\mu\,c} \, Y^{cb}+\omega^b_{\mu\,c} \, Y^{ac}\,.
\end{equation}
Fields of the tetrad and the connection can be related with one-forms
\begin{equation}
    e^a =e^a_\mu \, dx^\mu\,,\qquad 
\omega^{ab} =\omega^{ab}_\mu\,dx^\mu\,.
\end{equation}
All the geometric properties of the manifold can be expressed with these two variables and their exterior derivatives (contrary to metric formulation where one has to deal with second order derivatives; that's why it is often called the first order). Notice, that using to this p-forms allows us to hide some of the machinery underneath and simplify appearance of the expressions.\\\indent Just like in any other Yang-Mills theory, the object $\omega$ plays the role of the gauge potential for which we can give the field strength
$$R_{\mu\nu}{}^{ab}(\omega) =\partial_\mu \omega_\nu{}^{ab}-\partial_\nu \omega_\mu{}^{ab}+\omega_\mu{}^a{}_c\,\omega_\nu{}^{cb}-\omega_\nu{}^a{}_c\,\omega_\mu{}^{cb} \,,$$
used to give a rise to the curvature 2-form 
$$R^{ab}(\omega)=d\omega^{ab}+\omega^{a}{}_{c}\wedge \omega^{cb}=\frac{1}{2}R^{ab}_{\mu\nu}(\omega)\,dx^\mu\wedge dx^\nu\,.$$
To complete the whole description we introduce one more object: the torsion 2-form
\begin{equation}
    T^a_{\mu\nu}=T^{\rho}_{\mu\nu}\, e^a_\rho\,\qquad  T_{\mu\nu}{}^{a}=D^\omega_{\mu}\, e_{\nu}{}^{a}-D^\omega_{\nu}\, e_{\mu}{}^{a}\,,
\end{equation}
\begin{equation}
    T^{a}=D^\omega e^{a}=de^{a}+\omega^{a}{}_{b}\wedge e^{b}\,.
\end{equation}
The covariant derivative $D^\omega = (d + \omega)$ acts on them leading to two very important and useful Bianchi identities 
\begin{equation}
      D^\omega R^{ab}=0\,,\qquad\qquad   D^\omega T^a= R^{ab}\wedge e_b \,.            
\end{equation}
With building blocks: the fields ($\omega^{ab}, e^a$) and their field strengths ($R^{ab}, T^a$), we build Einstein-Cartan action with a negative cosmological constant ($\Lambda<0$) 
\begin{equation}\label{Einstein-Cartan}
    \displaystyle S=\frac{1}{64\pi G}~\int d^4x\,\epsilon^{abcd}(R_{\mu\nu\,ab}(\omega)\,e_{\rho\,k}\,e_{\sigma\,d}
-\frac{\Lambda}{3}e_{\mu\,a}e_{\nu\,b}e_{\rho\, c}e_{\sigma\, d})\epsilon^{\mu\nu\rho\sigma}\,.
\end{equation}
Written in \textit{form} language we rewrite it as
\begin{equation}\label{Palatini}
   S(\omega,e)=\frac{1}{ 32\pi G}\int \left(R^{ab}(\omega)\wedge e^{c}\wedge e^{d}-\frac{\Lambda}{6}\,  e^{a}\wedge e^{b}\wedge e^{c}\wedge e^{d}\right)\, \epsilon_{abcd}\,.
\end{equation}
Two independent fields return two field equations
\begin{equation}\label{Einstein equations}
    D(e^a\wedge e^b\,\epsilon_{abcd})=0\,,\qquad\qquad \left(R^{ab}(\omega)\wedge e^{c}-\frac{\Lambda}{3}\,  e^{a}\wedge e^{b}\wedge e^{c}\right)\epsilon_{abcd}=0\,.
\end{equation}
For an invertible tetrad the first one obviously means torsionless conditions 
\begin{equation}
   0= D^\omega e^a=d e^a+\omega^a{}_b\wedge e^b,
\end{equation}
which settles down the connection $\omega$ expressed as a function of the tetrad
\begin{equation}
\omega^{ab}_\mu(e)=e^{\nu\,a}\nabla_\mu
e^b_\nu=e^{\nu\,a}\left(\partial_\mu
e^b_\nu-\Gamma^\lambda_{\;\mu\nu}(g)e^{b}_\lambda \right)\,,
\end{equation} 
and one ends with equivalence between the tetrad and the metric formulation (where torsionless condition lies at its foundations). This relates the Riemann tensor with the curvature 2-form
\begin{equation}
   R^{\rho}{}_{\sigma\mu\nu}(g)=R^{ab}{}_{\mu\nu}(\omega(e))\,e_a^\rho \,e_{\sigma\,b}\,.
\end{equation}
One explicitly sees, that we do not follow the Einstein postulate, but the equations of motion determine vanishing of the torsion for the vacuum case (the same kind of feature is later repeated in the $BF$ theory). This situation will change in Chapter IV, where torsion won't vanish, forcing the connection to be composed from the spinor components.

Final form of (\ref{Palatini}) was achieved by the setting of the gauge Yang-Mills theory, however, it does not resemble its form. Yet, this could be achieved simply by replacing the Lorentz gauge group by the (A)dS group!

\section{MacDowell-Mansouri gravity}

This thesis focuses on applications of a model proposed by Freidel and Starodubtsev \cite{Freidel:2005ak}. Their construction has its roots in the work of Plebanski \cite{Plebanski:1977zz} and the procedure of MacDowell and Mansouri \cite{MacDowell:1977jt}. 

First concerns rewriting gravity as a topological $BF$ theory, where 'F' has to be understood as the curvature 2-form for the Lorentz gauge group (noted usually by $R^{ab}(\omega)$), contracted by the Killing form with $B^{cd}$ being an independent auxiliary field. One must also take into account a constraint, inserted to the action by the Lagrange multiplier, ensuring that on-shell $B_{ab}=\epsilon_{abcd}\, e^c \wedge e^d$. Solving this constraint immediately restores the Einstein-Cartan action
\begin{equation}\label{c2-1}
    S=\frac{1}{32\pi G}\int R^{ab}\wedge e^{c} \wedge e^d \, \epsilon_{abcd}\,.
\end{equation}
Purpose of this is that, since the $BF$ theory is topological (in a sense of lack of the dynamical degrees of freedom) its quantization is easier \cite{Plebanski:1977zz}, \cite{Broda:2005wk}, \cite{Mikovic:2005au}, therefore one might try to quantize General Relativity by rewriting it as the $BF$ action, and impose the constraints.

The latter approach of MacDowell and Mansouri combines the $so(1,3)$ \textit{spin connection} $\omega^{ab}$ and the \textit{tetrad} $e^a$, two independent variables in the Cartan theory, as the parts of the anti-de Sitter\footnote{Using de Sitter group is also possible, but because it has no applications to supergravity we will restrict ourselves only to the AdS case in the almost whole thesis.} $so(2,3)$ connection $A^{IJ}$ (notice, it does not mean expressing one by another)
\begin{equation}\label{c2-2}
A^{ab}_\mu=\omega^{ab}_\mu\,,\qquad  A^{a4}_\mu=\frac{1}{\ell}e^a_\mu\,.
\end{equation}
One can interpret this (see a review \cite{Wise:2006sm}) as a way of encoding the geometry of the spacetime $\mathcal M$ by parallel transport being "rolling" the anti-de Sitter manifold along the $\mathcal M$. Notice that the Lorentz part (with indices $a,b =0,1,2,3$) is embedded in the full symmetry group of the anti-de Sitter (for which $I,J = 0,1,2,3,4$) with Minkowski metric being $\eta_{IJ}=diag(-,+,+,+,-)$. To make dimensions right we need a length parameter, which has to be associated with a negative cosmological constant to recover standard General Relativity
\begin{equation}\label{c2-3}
    \frac{\Lambda}{3}=-\frac{1}{\ell^2}\,.
\end{equation} 
The $A^{IJ}$ can be further used to build curvature 2-form $ F^{IJ}(A)=\frac{1}{2} F^{IJ}_{\mu\nu}\,dx^\mu\wedge dx^\nu$
\begin{equation}\label{c2-4}
    F^{IJ}(A)=dA^{IJ}+A^{IK}\wedge A_K^{~~J}\,,\qquad F_{\mu\nu}^{IJ}=\partial_\mu A_\nu^{IJ} - \partial_\nu A_\mu^{IJ} +A_{\mu\,K}^I\, A_\nu^{KJ}-A_\nu^I{}_K\, A_\mu^{KJ}\,,
\end{equation}
which directly splits on the torsion
\begin{equation}\label{c2-5}
   F_{\mu\nu}^{a4} =\frac1{\ell}\left( \partial_\mu e_\nu^{a}  + \omega_\mu{}^a{}_b\, e_\nu^b -\partial_\nu e_\mu^{a}  - \omega_\nu{}^a{}_b\, e_\mu^b\right)=\frac1{\ell}\left( D^\omega_\mu e_\nu^{a}-D^\omega_\nu e_\mu^{a} \right)= \frac1{\ell}\, T_{\mu\nu}^a \,,
\end{equation}
and the so called AdS curvature
\begin{equation}\label{c2-6}
    F_{\mu\nu}^{ab}= R_{\mu\nu}^{ab}+ \frac1{\ell^2}\left(e_\mu^a\, e_\nu^b-e_\nu^a\, e_\mu^b \right)\, ,
\end{equation}
with the standard Lorentz curvature
\begin{equation}\label{c2-7}
    R_{\mu\nu}^{ab}=\partial_\mu\omega_\nu^{ab} - \partial_\nu\omega_\mu^{ab} +\omega_{\mu\,c}^a\, \omega_\nu^{cb}-\omega_\nu^a{}_c\, \omega_\mu^{cb}\,.
\end{equation}
One cannot use the full curvature $F^{IJ}(A)$ for building an action of the dynamical theory in four dimensions. However, such a formulation is possible with the use of the group dual ($\star$) and breaking the gauge symmetry by projecting full curvature down to Lorentz indices
\begin{equation}
    F^{IJ}\quad\rightarrow\quad \hat{F}^{IJ}= F^{ab}\quad\quad \mathrm{where}\quad\quad F^{ab}=R^{ab}+\frac{1}{\ell^2}e^a\wedge e^b\,.
\end{equation} 
General Relativity seen as a gauge symmetry breaking theory then emerges from the action
\begin{equation}\label{c2-8}
    S_{MM}(A)=\frac{\ell^2}{64\pi G} \int tr\big(\hat F \wedge \star \hat F\big)\,,
\end{equation}
\begin{equation}\label{c2-9}
    S_{MM}(A)=\frac{\ell^2}{64\pi G} \int \left(R^{ab}+\frac{1}{\ell^2}e^a\wedge e^b\right)\wedge \left(R^{cd}+\frac{1}{\ell^2}e^c\wedge e^d\right)\epsilon_{abcd}\,.
\end{equation}
This intriguing theory of gravity, based on the AdS/dS group, and underneath reducing to
\begin{equation}\label{c2-10}
  32\pi G\, S_{MM}=\int \mathrm{Einstein/Cartan}+\frac{1}{\,2\ell^2}\int  \mathrm{cosmological}+\frac{\ell^2}{2}\int \mathrm{Euler} \,, 
\end{equation}
ties together a cosmological constant and the Euler invariant (term quadratic in curvature forms). At the same time, from a formal point of view, it established a form known from Yang-Mills theories.

The field equations remain the same as in the Einstein-Cartan case (\ref{Einstein equations}), because the variation concerning an Euler term leads to $D^\omega R^{ab}$, which vanishes due to the Bianchi identity.

It should be noticed that in this setting the difference between General Relativity and Yang-Mills theory is reduced to a constraint (achieved by breaking full symmetry), which has an extremely simple interpretation: it is the torsion-free condition coming from the lack of fermionic content \cite{BottaCantcheff:2000ti}. 

Presence of fermions changes the Einstein field equations according to eq. (\ref{ch2-6}), where a spin couples to the torsion. One can try to use prescription given by MacDowell and Mansouri to combine the field equations now expressed in the \textit{p-form} language
\begin{eqnarray}
 \frac{1}{2}\epsilon_{abcd} (R^{ab}+\frac{1}{\ell^2} e^a\wedge e^b)\wedge e^c &=&16\pi G\, t_d\\
\epsilon_{abcd}\, T^c\wedge e^d&=& 16\pi G \,s_{ab}
\end{eqnarray}
with $t_d$ being energy-momentum density 3-form, and $s_{ab}$ being 3-form of spin density.\\
They can be evaluated in the tetrad basis, with corresponding tensors in $T_x$:
\begin{equation}
    t_d=\frac{1}{3!}T_{df}\,\epsilon^{abcf} \,e_a\wedge e_b\wedge e_c\,,\qquad 
s_{mn} =\frac{1}{3!}S^f{}_{mn}\,\epsilon^{abcf}\, e_a\wedge e_b\wedge e_c\,.
\end{equation}
It is easy from here to show that it could be generalized as
\begin{equation}
(\epsilon_{IJKLa}  F^{KL}\wedge A^{a4} - 16\pi G\, s_{IJ})\wedge\delta A^{IJ}=0\,
\end{equation}
\begin{equation}
 \delta (\epsilon_{abcd4} F^{ab}\wedge F^{cd}) - 16\pi G\, s_{IJ}\wedge\delta A^{IJ}=0\,,
\end{equation}
This naturally unites the forms of the energy-momentum and spin density \footnote{For the description of BF theory trying to reconcile with the point particles with mass and spin one should check \cite{KowalskiGlikman:2008fj}.} into single object
\begin{equation}
    s_{IJ}=\left\{
\begin{array}{l l}
s_{ab}\\
s_{a4}=t_a \,.   
\end{array}\right.
\end{equation}

In above we have used covariant derivative $D^A$ acting on $\epsilon_{abcd}=\epsilon_{abcd4}$ which is not invariant tensor for the $SO(2,3)$, therefore $D^A \epsilon_{abcd4}=\epsilon_{IJKLm} A^{m}{}_{4}$.\newline

It is quite remarkable, that besides reproducing gravity from this scheme, authors of \cite{MacDowell:1977jt} were also able to use it to provide the $\mathcal{N}=1$ supergravity action by extending the definition of the connection to cover the spin $3/2$ gravitino field. This will be explored in more details in Chapter IV. At this point we also postpone further motivation and the rest of features offered by the action $S_{MM}(A)$ to Chapter V, where we will see how essential it is for a proper variation principle and finite values of the black hole's mass and the angular momentum in the AdS asymptotic spacetimes. 

Let's now turn to presenting $BF$ model sharing main advantages of the construction above and proving great usefulness in the discussion concerning the Immirzi parameter, and the rest of topological terms.

\chapter{Deformed BF theory as a theory of gravity}

\section{Deformed topological BF theory}

Quite recently MacDowell-Mansouri formalism was generalized to the form of a deformation of the topological $BF$ theory based on the (anti) de Sitter gauge group. Such a construction of gravity has been developed by Smolin, Freidel and Starodubtsev \cite{Freidel:2005ak},  \cite{Wise:2006sm}, \cite{Smolin:2003qu}, \cite{Starodubtsev:2003xq}. Instead of directly introducing the constraints by the Lagrange multipliers it was suggested to use a term built in the same fashion like in the MacDowell-Mansouri model. The action, apart of the 1-form $so(2,3)$-valued connection $A^{IJ}$ and built from it curvature $F^{IJ}(A)$, should be appended with auxiliary $so(2,3)$-valued 2-form $B^{IJ}$ field
\begin{equation}\label{equation BF}
 16\pi \, S_{BF}(A,B)= \int tr \left( B\wedge F - \frac{\alpha}{4}\hat B\wedge \star \hat B \right)\,,
\end{equation}
By trace in the first term we understand the Killing form for full AdS group, and in the second for its Lorentz subgroup.

After solving the equations coming from the variations
$$\delta B_{a5}:~F^{a5}=\frac{1}{\ell}T^a=0,\qquad\qquad\qquad \delta B_{ab}:~ F^{ab}=\frac{\alpha}{2}\epsilon^{abcd} B_{cd}$$ and plugging them back into the action, for $\alpha$ equals $\frac{G\Lambda}{3}$, we achieve the equivalence 
$$S_{BF}(A,B)\equiv S_{MM}(A)\,.$$

MacDowell-Mansouri proposal was the construct for itself. Now in the $BF$ theory context it becomes the scheme to build a structure on the topological vacuum, understood as a part of theory without the degrees of freedom. With the form of (\ref{equation BF}) we are facing very interesting reformulation having an appearance of some new kind of perturbation theory, in which General Relativity is reproduced as a first order perturbation around the topological vacuum. Symmetry breaking occurs in the last term with dimensionless coefficient proportional, for the observed de Sitter space, to extremely small parameter $\alpha=G\Lambda/3 \sim 10^{-120}$ \cite{Freidel:2005ak}. 

Such a $BF$ model can be extended, not only to reproduce MacDowell-Mansouri action, but also to incorporate the Immirzi parameter, as well as another (Pontryagin and Nieh-Yan) topological terms. Before we go any further let's explain why we want to include them.

\section{Immirzi parameter}
Program of Loop Quantum Gravity started with the discovery of the Ashtekar variables: SL(2,C) (anti) selfdual connection \cite{Ashtekar:1986yd} and its conjugated momentum\footnote{Here group indices are restricted to $a,b=1,2,3$, and spacetime indices to spacelike $i,j=1,2,3$.}
$$\displaystyle\omega_i^a = \omega^{0a}_i+ \frac{i}{2}\epsilon^{0abc}\omega_{i\,bc}, \qquad\mathcal{P}_{a}^i=\frac{4}{16\pi G}\epsilon_{abc}\epsilon^{ijk}e^b_j\,e^c_k\,.$$
Phase space of gravity described by the self/antiselfdual connections leads to significant simplification of the General Realtivity Hamiltonian. However this means that have to deal with a complex formulation, so the reality conditions has to be imposed. Because they are hard to implement, Barbero and Immirzi \cite{Barbero:1994ap, Immirzi:1996di}, independently,  suggested to use real parameter, usually denoted as $\gamma$, to replace imaginary unit. For a phase space of gravity described by
$$\displaystyle{}^\gamma\omega_i^a = \omega^{0a}_i+ \frac{\gamma }{2}\epsilon^{0abc}\omega_{i\,bc}\,,\qquad\mbox{and} \qquad\mathcal{P}_{a}^i=\frac{4}{16\pi G}\epsilon_{abc}\epsilon^{ijk}e^b_j\,e^c_k,$$ we can see that the Poisson bracket is given by
$$\displaystyle \{{}^\gamma\omega_i^a(x), \mathcal P_b^j (y) \}=\gamma \delta(x-y)\delta_{i}^j\delta^{a}_b\,.$$
The Immirzi-Barbero parameter came with a price, which was paid in loosing some of the earlier achieved simplicity of the constraints building a Hamiltonian for GR, but it still gives some hopes for the progress of the canonical gravity program.

One can show that formulation leading to the Ashtekar variables is just the (anti)selfdual Einstein-Cartan action. Now, in case of the Barbero-Immirzi variables, the usual first order Lagrangian will be accompanied with a term (where again we have full $a,b=0,1,2,3$) 
\begin{equation}\label{Holst}
    \frac2{64\pi G\gamma}\, \epsilon^{\mu\nu\rho\sigma}R_{\mu\nu}{}^{ab}\, e_{\rho\,a}\,e_{\sigma\, b}\, ,
\end{equation}
in literature known as the Holst term \cite{Holst:1995pc}.

The presence of $\gamma$ parameter in the gravity Lagrangian was for many years overlooked because the corresponding Holst term by virtue of the second Bianchi identity does not contribute to the equations of motion on shell, when torsion vanishes, but it should be stressed that it is not topological and it influences the canonical structure of the theory. In quantum theory $\gamma$ might be relevant, because it controls the rate of quantum fluctuations of torsion. 

For the discussion concerning coupling the torsion to the matter fields with spin-1/2, and possibility of the physical effects of the Immirzi parameter see \cite{Perez:2005pm} and \cite{Freidel:2005sn} with the counter arguments presented in \cite{Kaul:2007gz}.

In spite of the fact that Immirzi parameter is not visible in vacuum field equations, its presence leads to modifications of the phase space structure of the theory, which in turn make it reappear in the spectra of Loop Quantum Gravity area and volume operators (see e.g., \cite{Rovelli:2004tv}, \cite{Thiemann:2007zz}) and in the calculation of black hole entropy. This subject will be analyzed closer in the context of R. Wald's approach in Chapter IV, whereas in Chapter V we will try to answer to a question if the Immirzi parameter can be coupled to the spin-3/2 fields, and modify $\mathcal{N}=1$ supergravity.

\section{Topological terms}

The Wilsonian perspective, telling that one should include in the action all terms that can be constructed from the fields and are compatible with the symmetries of the theory, is a powerful guiding principle in constructing theories with the given field content and symmetries. Every possible term would come in the action with its own coupling constant, and one could ask if there is an additional principle that could be used to reduce
the number of independent parameters of the theory. This chapter shows that
it can be achieved in the $BF$ framework, which is effectively governed only by the three constants: $\gamma$, $G$, and $\Lambda$ (see eq. (\ref{ch3-1b})).

In the context of first order gravity we have to deal with two fields, tetrad $e^a$ and connection $\omega^{ab}$, and two symmetries, local Lorentz invariance and spacetime diffeomorphisms. If we implement
the diffeomorphism invariance, assuming that the action of gravity
is written as a four form polynomial constructed from the tetrad and the connection, the
list of possible terms turns out to be rather short and includes:
\begin{itemize}
\item Einstein-Cartan action
\begin{equation}\label{action-1}
  \mathcal{L}_{EC}= R^{ab}\wedge e^{c}\wedge e^{d}\,\epsilon_{abcd}\, ,
\end{equation}
\item Cosmological term
\begin{equation}\label{action-2}
    \mathcal{L}_{\Lambda}=e^{a}\wedge e^{b}\wedge e^{c}\wedge e^{d}\, \epsilon_{abcd}\, ,
\end{equation}
\item Holst term
\begin{equation}\label{action-3}
 H_4 =  R^{ab}\wedge e_a \wedge e_b\, ,
\end{equation}
\item topological Pontryagin, Euler and Nieh-Yan terms
\begin{eqnarray}\label{action-4}
\mathcal{E}_4 &=& R^{ab}\wedge R^{cd} \,\epsilon_{abcd}\, ,\nonumber\\
\mathcal{P}_4 &=& R^{ab}\wedge R_{ab}\, ,\nonumber\\
\mathcal{NY}_4&=& T^a \wedge T_a - R^{ab}\wedge e_a \wedge e_b\, .
\end{eqnarray}
\end{itemize}
Pontryagin class \cite{Chandia:1997hu}
is related to the Chern-Simons class, whereas in the case of the tangent bundle of a smooth manifold, the Euler class generalizes classical notion of Euler characteristic $\chi(\mathcal{M})$. Four dimensional Euler term is also an equivalent of the Gauss-Bonnet term (being a part of Lovelock and Lanczos gravity series), which in MacDowell-Mansouri gravity comes with a fixed weight associated with a cosmological constant. Nieh-Yan class \cite{Nieh:1981ww} is nothing else than the difference between Pontryagin terms for the full AdS $SO(2,3)$-connection $A^{IJ}$ and the Lorentz $SO(1,3)$-spin connection $\omega^{ab}$ (for its applications see \cite{Chandia:1997hu}, and \cite{Mercuri:2007ki})
\begin{equation}
    F^{IJ}(A)\wedge F_{IJ}(A)=R^{ab}(\omega)\wedge R_{ab}(\omega)-\frac{2}{\ell^2}\left(T^a\wedge T_a - R_{ab}\wedge e^a \wedge e^b \right)
\end{equation}
\begin{equation}
    \mathcal{P}_5(A)=\mathcal{P}_4(\omega)-\frac{2}{\ell^2}\mathcal{NY}_4
\end{equation}
All of these terms
\begin{eqnarray}\label{topological terms}
& &  P_4=\frac{1}{4}\int d^4x\, R_{\mu\nu\,ab}R^{ab}_{\rho\sigma}\,\epsilon^{\mu\nu\rho\sigma}\,,\nonumber\\
& &E_4=\frac{1}{4}\int d^4x\,  R_{\mu\nu\,ab}R_{\rho\sigma\,cd}\epsilon^{abcd}\,\epsilon^{\mu\nu\rho\sigma}\,,\nonumber\\
& & NY_4=\frac{1}{4}\int d^4x\,  (T_{\mu\nu\,a}T^{a}_{\rho\sigma}-2\,R_{\mu\nu\,ab}e_{\nu}^{\,a}e_{\rho}^{\, b})\,\epsilon^{\mu\nu\rho\sigma}\,,
\end{eqnarray}
quite remarkably, could be written as the total derivatives, thus, in fact they can't influence the bulk dynamics. 

\noindent With the second Bianchi identity it is straightforward to check that
\begin{align}
 NY_4&=2\int \partial_\mu \Big( e_{\nu\;\alpha} T^\alpha_{\rho\sigma} \Big)\,\epsilon^{\mu\nu\rho\sigma}=4\int \partial_\mu \Big( e_{\nu\;\alpha} \mathcal D^\omega_\rho e^{\;\alpha}_{\sigma} \Big)\,\epsilon^{\mu\nu\rho\sigma}\nonumber\\
 P_4&=\int R_{\mu\nu\,ab}R^{ab}_{\rho\sigma}\,\epsilon^{\mu\nu\rho\sigma}=4\int\partial_\mu C^\mu(\omega)\label{almost-chern}
\end{align}
where $C^\mu$ is expressed by the Chern-Simons term (a subject of vast research in 2+1 gravity)
\begin{equation}
   C^\mu(\omega)= \Big(\omega_{\nu\;ab}\,\partial_\rho \omega_\sigma^{ab}+\frac{2}{3}\omega_{\nu\;ab} \,\omega_{\rho\;\,c}^{\;a} \,\omega_{\sigma}^{cb}\Big)\,\epsilon^{\mu\nu\rho\sigma}\,,
\end{equation}
allowing us (using generalization $\omega^{ab}\to A^{IJ}$) to write the relations (\ref{almost-chern}) as
\begin{align}
 C^{\mu}(A)=C^{\mu}(\omega)-\frac{2}{\ell^2}(e_{a \nu}D^{\omega}_\rho e_\sigma^a)\epsilon^{\mu\nu\rho\sigma}\,.
\end{align}
To apply the same to the Euler term it is necessary to make transition to the self- and anti-selfdual connections ${}^{\pm}\omega$ and associate it with their curvatures
\begin{equation}
    {}^{\pm}\omega^{ab}_{\mu}=\frac{1}{2}\Big( \omega_{\mu}^{ab}\mp\frac{i}{2}\epsilon^{ab}_{\;\;cd }\omega_{\mu}^{cd} \Big)
\end{equation}
\begin{equation}
    {}^{\pm} R^{ab}_{\mu\nu}=\frac{1}{2}\Big(\frac{1}{2}\delta^{ab}_{\;cd} \mp\frac{i}{2}\epsilon^{ab}_{\;\;cd} \Big) R_{\mu\nu}^{cd}\, ,\quad {}^{\pm}R^{ab}_{\mu\nu}=\frac{1}{2}\Big( R_{\mu\nu}^{ab}\mp\frac{i}{2}\epsilon^{ab}_{\;\;cd }R_{\mu\nu}^{cd} \Big)\,,
\end{equation}
which allows us to write 
\begin{equation}
  \epsilon^{\mu\nu\sigma\rho}\,  {}^{\pm} R^{ab}_{\mu\nu} {}^{\pm} R_{\rho\sigma\;ab}=
\frac{1}{4}\epsilon^{\mu\nu\sigma\rho}\, \Big(  2R^{ab}_{\mu\nu}\, R_{\rho\sigma\;ab} \mp i  R^{ab}_{\mu\nu} \, R_{\rho\sigma}^{cd}\, \epsilon_{abcd}\Big)\,.
\end{equation}
\begin{equation}
 4\partial_\mu \mathcal{C}^\mu({}^\pm\omega)=
\frac{1}{4}\Big(  2\mathcal{P}_4(\omega) \mp i\mathcal{E}_4(\omega)\Big)\,.
\end{equation}
Such combinations bring the final expressions
\begin{align}
P_4&=4\int \Big(\partial_\mu \mathcal C ^\mu (^+\omega)+ \partial_\mu  \mathcal C^\mu (^-\omega) \Big)\\
E_4&=8i \int\Big(\partial_\mu  \mathcal C ^\mu (^+\omega)-\partial_\mu  \mathcal C^\mu (^-\omega)\Big)\,,
\end{align}
which, although having imaginary unit inside, stay real.

Euler term has proved its worth completing the Einstein-Cartan action, and allowing for rewriting it as the YM theory. Although it is topological, and does not change (at least classically) the equations of motion of gravity, it will be essential for the gravitational Noether charges. Other terms will exhibit similar usefulness being complementary to the Holst term.
\section{Freidel-Starodubtsev BF model}
Resulting from the Barbero-Immirzi variables the Holst modification \cite{Holst:1995pc} can be easily incorporated in the action (\ref{equation BF}) just by additional term being quadratic in $B$ fields, which brings the rest of possible terms. The action proposed by Freidel and Starodubtsev \cite{Freidel:2005ak}: 
\begin{equation}\label{ch3-1}
 16\pi \, S(A,B)= \int F^{IJ}\wedge B_{IJ} -\frac{\beta}{2} B^{IJ}\wedge B_{IJ} - \frac{\alpha}{4}\epsilon^{abcd4} B_{ab}\wedge B_{cd}
\end{equation}
yields the desired extension including the Immirzi parameter 
expressed as $\gamma=\frac{\beta}{\alpha}$, and dimensionless parameters $\alpha$, $\beta$ related to the gravitational and cosmological constants
\begin{equation}\label{ch3-1b}
     \alpha= \frac{G\Lambda}{3\,(1+\gamma^2)}\,,\qquad
\beta= \frac{\gamma G\Lambda}{3\,(1+\gamma^2)}\quad \mathrm{with}\quad \Lambda=-\frac3{\ell^2}\,.
\end{equation}
The first two terms in the action above are invariant under the action of
local $so(2,3)$ gauge symmetries if $B^{IJ}$ transform under
these symmetries like curvatures. The third term,
however, is invariant only under the action of a subgroup of the
Anti de Sitter group, leaving $\epsilon^{IJKL4}$
invariant\footnote{The totally antisymmetric
symbol $\epsilon^{IJKLM}$, defined by $ \epsilon^{01234}=1$, is an
invariant tensor of the algebra $so(2,3)$. With one of the direction fixed we define $\epsilon^{abcd4}=\epsilon^{abcd}$, being an invariant tensor of the algebra $so(1,3)$.}, which is Lorentz subgroup with the algebra $so(1,3)$. This term can be thought of as a constraint (that's why it is often called as the constrained $BF$ model), explicitly breaking the local
translational invariance and rendering the action only local-Lorentz
invariant (for an extensive review see \cite{Wise:2006sm}, and new \cite{Freidel:2012xp}).

To see that (\ref{ch3-1}) is equivalent to the action of general relativity we solve it for $B_{\mu\nu}^{IJ}$ and substitute the result back to the Lagrangian. One finds then
\begin{equation}\label{ch3-2}
 F_{}{}^{a4}=\beta    B_{}{}^{a4}\,,
\end{equation}
\begin{equation}\label{ch3-3}
   F_{}{}^{ab}=\beta B_{}{}^{ab}+\frac{\alpha}{2}\, \epsilon^{abcd}\, B_{}{}_{cd},
\end{equation}
with its inverse
\begin{equation}\label{ch3-4}
B_{}{}^{ab}=\frac{1}{\alpha^2+\beta^2}\left(\beta F_{}{}^{ab}-\frac{\alpha}2\, \epsilon^{abcd}\, F_{}{}_{cd}\right)\, .
\end{equation}
Through solving these field equations for the $B$ fields
we express the resulting Lagrangian in terms of the
$so(1,3)$-connection $\omega$, and the tetrad $e$ in quite compact form, which we will later find particularly convenient
\begin{equation}\label{ch3-5}
 S(\omega,e)=\frac{1}{16\pi} \int\left( \frac{1}{4}M^{abcd} F_{ab}\wedge F_{cd}-\frac{1}{\beta\ell^2} \,T^a \wedge T_a\right)\, 
\end{equation}
with
\begin{equation}\label{ch3-6}
M^{ab}{}_{cd}=\frac{\alpha}{(\alpha^2+\beta^2)}( \gamma\, \delta^{ab}_{cd}-\epsilon^{ab}_{\;\;cd}) \equiv -\frac{\ell^2}{G}( \gamma\, \delta^{ab}_{cd}-\epsilon^{ab}_{\;\;cd}) \,.   
\end{equation}
Remarkably, such action written explicitly includes all six possible terms of tetrad gravity in four dimensions, fulfilling all the necessary symmetries, and is governed only by $G, \Lambda$, and $\gamma$ 
\begin{eqnarray}\label{ch3-7}
 32\pi G\, S&=&\int R^{ab}\wedge e^{c}\wedge e^{d}\,\epsilon_{abcd}+
\frac{1}{\,2\ell^2}\int  e^{a}\wedge e^{b}\wedge e^{c}\wedge e^{d}\, \epsilon_{abcd}+\frac{2}{\gamma}\int R^{ab}\wedge e_{a}\wedge e_{b}\nonumber\\
&+&\frac{\ell^2}{2}\int R^{ab}\wedge R^{cd} \,\epsilon_{abcd} -\ell^2\gamma\int R^{ab}\wedge R_{ab}\nonumber\\
&+&\frac{\gamma^2+1}{\gamma}\int 2\,(T^a \wedge T_a - R^{ab}\wedge e_a \wedge e_b)\,.
\end{eqnarray}
With all the spacetime indices written down, it reads as follows
\begin{eqnarray}\label{ch3-8}
64 \pi G\, S&=&\int\epsilon^{abcd}(R_{\mu\nu\,ab}e_{\rho\,c}e_{\sigma\,d}
-\frac{\Lambda}{3}e_{\mu\,a}e_{\nu\,b}e_{\rho\, c}e_{\sigma\, d})\epsilon^{\mu\nu\rho\sigma}+\frac{2}{\gamma}\int R_{\mu\nu\,ab}\,e_{\nu}^{\,a}e_{\rho}^{\, b}\,\epsilon^{\mu\nu\rho\sigma}\nonumber\\
&+&\frac{\gamma^2+1}{\gamma}NY_4
+\frac{3\gamma}{2\Lambda}P_4-\frac{3}{4\Lambda}
E_4\, .
\end{eqnarray}
One can see that structure standing behind analyzed $BF$ model turns out to be the combination of the Cartan--Einstein action (\ref{Einstein-Cartan}) with a cosmological constant term and the Holst term (\ref{Holst}), accompanied by the topological Euler, Pontryagin and Nieh-Yan terms (\ref{topological terms}).

Field equations resulting from (\ref{ch3-1}) are effectively the standard
vacuum Einstein equations. The field equations on the level of $BF$ theory read
\begin{equation}\label{ch3-9}
 (D^A_{}\, B_{})^{IJ}=0\, ,
\end{equation}
\begin{equation}\label{ch3-10}
F_{}{}^{IJ} -\beta\, B_{}{}^{IJ}-\frac\alpha2\, \epsilon^{IJKL4}\, B_{}{}_{KL}=0\, ,
\end{equation}
where $D^A_{}$ is the covariant derivative defined by the connection $A^{IJ}$, so that
\begin{equation}\label{ch3-11}
    (D^A_{}\, B_{})^{IJ}=d B_{}{}^{IJ} +A_{}^I{}_K\wedge B_{}{}^{KJ}+ A_{}^J{}_K\wedge B_{}{}^{IK}\, .
\end{equation}
Using decomposition on the spin connection and tetrad we rewrite (\ref{ch3-9}) as
\begin{equation}\label{ch3-12}
D^\omega B^{ab} + \frac1\ell\, e^ a\wedge B^{b4}-\frac1\ell\, e^b\wedge B^{a4}=0\, ,
\end{equation}
\begin{equation}\label{ch3-13}
D^\omega B^{a4} - \frac1\ell\, e_b\wedge B^{ab}=0\, ,
\end{equation}
where we can use relation (\ref{ch3-10}) to get rid off all $B$ fields. One can also extract field equations directly from the action (\ref{ch3-5}) expressed solely in terms of the tetrads and the connections. Any way, although in starting action we had the torsion, now, just like in the first order formalism, the equations of motion determine its vanishing. Finally from the variation principle we have
\begin{equation}\label{ch3-14}
(\frac{1}{\gamma}\,
\delta^{ab}{}_{cd}+\epsilon^{ab}{}_{cd})\, F_{ab}\wedge e^c=0\,,\qquad
D^\omega \left((\frac{1}{\gamma}\,
\delta^{ab}{}_{cd}+\epsilon^{ab}{}_{cd})\,e_a \wedge e_b\right)=0\,.
\end{equation}
For $\gamma^2\neq -1$, and invertible tetrad, the second condition means vanishing of the torsion $T^a=D^\omega e^a=0$, thus the first term reduces to the standard Einstein field equations:
\begin{equation}\label{ch3-15}
    \left(R^{ab}\wedge e^c+\frac{1}{\ell^2}e^a\wedge e^b\wedge e^c \right)\,\epsilon_{abcd}=0\,.
\end{equation}
Equations are unaffected by the quadratic in curvatures Euler and Pontryagin terms because their variation leads to $D^\omega R^{ab}$, which vanishes due to the Bianchi identity, where the Nieh-Yan contributes to equations in the Immirzi terms, which vanish anyway by the virtue of the second Bianchi identity $R^{ab}\wedge e_b=D^\omega T^a$ and vanishing of a torsion. 

There are many advantages of such formulation of gravity. As was stressed in \cite{Freidel:2005ak}, on the level of $BF$ theory it makes the kinetic term of the Lagrangian quadratic in fields, which makes the standard methods of quantum field theory applicable, contrary to the case of the tetrad formalism, in which the kinetic term is trilinear. Lagrangian (\ref{ch3-1}) contains two groups of terms. First we have the terms describing a topological field theory of $BF$ type for the gauge group, which is chosen to be the de Sitter $SO(1,4)$ or anti-de Sitter $SO(2,3)$ group. These terms generate the topological vacuum of the theory. The remaining term is responsible for the dynamics of gravity, and is chosen in such a way so as to break the topological theory gauge symmetry down to the local $SO(1,3)$ Lorentz symmetry of gravity.

It opens an exciting possibility of a manifestly diffeomorphism invariance perturbative approach to quantum gravity (with and without matter sources) \cite{Freidel:2005ak}, \cite{KowalskiGlikman:2006mu}, \cite{KowalskiGlikman:2008fj}, in which the gauge breaking term is regarded as a perturbation around topological vacuum described by $BF$ theory. This approach introduces the Immirzi parameter $\gamma$ to the theory in a natural way. One should also point that many various calculations are much simpler than in the case of explicit tetrad gravity. 

This model shares main features of the MacDowell-Mansouri proposal and goes beyond, as we will see by turning to its applications. It could be used as a tool to analyze the wide range of topics including supergravity, black hole thermodynamics, AdS-Maxwell algebra, and canonical analysis. In this task we will be interested in the formal side of the constructions, as well as the relevance of the Immirzi parameter and the topological terms in any of \mbox{these topics.}

\part{Main results}

\chapter{Super-BF theory}

In this chapter we extend the construction of the $BF$ theory to ${\mathcal N}=1$ supergravity, generalizing the results reported in \cite{Townsend:1977qa} to the presence of the Immirzi parameter. It turns out that the Holst term (\ref{Holst}) is replaced by its supersymmetrized counterpart, which effectively does not influence supergravity. 

\section{Supergravity and BF theory}

Supersymmetry is a powerful idea based on irresistible beauty of the symmetry relating fermions with bosons. Here we are going to touch only small part of this subject, namely, focus on supergravity part arising from combining a theory of gravity with principles of supersymmetry. It means introducing the gravitino, spin 3/2 field, in addition to the graviton represented by the spin connection and the tetrad. Crucial works \cite{Townsend:1977qa}, \cite{Ferrara:1976kg}, and \cite{VanNieuwenhuizen:1981ae} showed that this formulation is in fact possible, revealing that antisymmetric part of the connection is not longer obsolete formal generalization, but something much deeper. One finds that this theory contains antisymmetric part of a connection related to a gravitino, proving that $\omega$ might be other than just Riemannian one.

Achieved in the late 70's the action for supergravity (SUGRA) \cite{Ferrara:1976kg} was a result of inserting by hand necessary terms to make the final Lagrangian supersymmetric, and restore  the Rarita--Schwinger equations. MacDowell with Mansouri tried to find some other premise with more meaningful setting, and the goal was achieved in their formulation \cite{MacDowell:1977jt} with crucial role of $OSp(1,4)$ superalgebra replacing the anti-de Sitter algebra, and extending definition of the connection $\mathbb{A}$ to contain not only $\omega^{ab}$ with $e^a$, but also the Majorana spinor field $\psi$. We are going to repeat it in the supersymmetric extension of the $BF$ theory to ${\cal N}=1$ supergravity, which will help us understand such a proposal in more details.

\section{Gauging the superalgebra}

Let us briefly recall how MacDowell and Mansouri scheme works, which will be quite relevant for later investigation of the AdS modification in Chapter VI. The starting gauge algebra is $SO(2,3)$ with the anti de Sitter generators $\mathcal{M}_{IJ}$ following the commutation rule
\begin{equation}\label{c3-1}
 [   \mathcal{M}_{IJ},\mathcal{M}_{KL}]=-i(\eta_{IK}\mathcal{M}_{JL}+\eta_{JL}\mathcal{M}_{IK}-\eta_{IL}\mathcal{M}_{JK}-\eta_{JK}\mathcal{M}_{IL})\,,
\end{equation}
and the metric tensor $\eta_{IJ}$ (for $I,J=0,\ldots,4$) having the
signature $(-,+,+,+,-)$. After decomposing generators $\mathcal{M}_{IJ}$ into Lorentz $SO(1,3)$ $\mathcal{M}_{ab}$ and translation $\mathcal{P}_a =\mathcal{M}_{a4}$ we find
\begin{equation}\label{c3-2}
[\mathcal{M}_{ab},\mathcal{M}_{cd}]=-i(\eta_{ac}\mathcal{M}_{bd}+\eta_{bd}\mathcal{M}_{ac}-\eta_{ad}\mathcal{M}_{bc}-\eta_{bc}\mathcal{M}_{ad})\,,
\end{equation}
\begin{equation}\label{c3-3}
[\mathcal{M}_{ab},\mathcal{P}_c]=-i(\eta_{ac}\mathcal{P}_{b}-\eta_{bc}\mathcal{P}_{a})\, ,\qquad
[\mathcal{P}_{a},\mathcal{P}_{b}]=-i\eta_{44}\mathcal{M}_{ab}=i\mathcal{M}_{ab}\, .
\end{equation}
Therefore, this algebra splits into its Lorentz and translational parts, generated by $\mathcal{M}_{ab}$ and $\mathcal{P}_a=\mathcal{M}_{a4}$ generators, respectively. Accordingly we can split the gauge field
\begin{equation}\label{c3-4}
\mathbb{A}_{\mu}=\frac{1}{2}A_\mu{}^{IJ}\mathcal{M}_{IJ}=\frac{1}{2}\omega_\mu{}^{ab}
\mathcal{M}_{ab}+\frac{1}{\ell}e_\mu{}^{a}
\mathcal{P}_{a}
\end{equation}
with $\omega^{ab}_\mu$ being the Lorentz connection and $e_\mu^{a}$ identified with the tetrad. 

We know that for the gauge field $\mathbb{A_{\mu}}$ we can build the curvature
\begin{equation}\label{c3-5}
 \mathbb{F}_{\mu\nu}(\mathbb{A})=\partial_\mu\mathbb{A}_{\nu}-\partial_\nu\mathbb{A}_{\mu}-i[\mathbb{A}_{\mu},\mathbb{A}_{\nu}]\, ,
\end{equation}
which, with the help of commutators above, can be decomposed into translational and Lorentz parts defined by the torsion (\ref{c2-5}) and the curvature (\ref{c2-6}).

Used for the purpose of supersymmetry the superalgebra $OSp(1,4)$ of course contains the bosonic part being the $SO(2,3)$ algebra. To go further we introduce $\gamma$-matrices satisfying the standard Clifford algebra
\begin{equation}\label{c3-6}
    \{\gamma^a, \gamma^b\}=2\eta^{ab}, \quad \eta^{ab}=\mbox{diag}(-,+,+,+)\, ,
\end{equation}
with
$$
\gamma^{5} =  \left(
\begin{array}{cc}
-i\sigma^2 & 0\\
0 & i\sigma^2
\end{array}
\right)
\qquad \mbox{and}\qquad
\gamma^{0} = \left(
\begin{array}{cc}
0 & -i\sigma^2\\
-i\sigma^2 & 0
\end{array}
\right)
$$
\begin{equation}\label{c3-7}
 \gamma^{1} = \left(
\begin{array}{cc}
\sigma^3 & 0\\
0 & \sigma^3
\end{array}
\right)
\quad
\gamma^{2} = \left(
\begin{array}{cc}
0 & i\sigma^2\\
-i\sigma^2 & 0
\end{array}
\right)
\quad
\gamma^{3} =  \left(
\begin{array}{cc}
-\sigma^1 & 0\\
0 & -\sigma^1
\end{array}
\right)\,.
\end{equation}
One checks that the following combinations of $\gamma$ matrices
$$
    m_{a4}=\frac{1}2\, \gamma_a,\quad m^{a4}=-\frac{1}2\, \gamma^a, \quad m_{ab}=m^{ab}=\frac14\, [\gamma_a, \gamma_b]=\frac12 \gamma_{ab}
$$
forms a representation of the $SO(2,3)$ through $\mathcal{M}_{IJ}=im_{IJ}$.\\ 
The supersymmetry generator $Q$ transforms as a (Majorana) spinor with respect to $SO(2,3)$
\begin{equation}\label{c3-9}
    [M_{IJ},Q_\alpha]=-i(m_{IJ})_{\alpha}^{~~\beta}\, Q_\beta \, ,\quad\mathrm{i.e.}
\quad [M_{ab}, Q] =-\frac{i}2\, \gamma_{ab}\, Q, \quad [P_a, Q] = -\frac{i}{2}\, \gamma_a\, Q\,.
\end{equation}
Finally the anicommutator of two supersymmetry generators reads
\begin{equation}\label{c3-10}
\{Q_\alpha, Q_\beta\}=-im^{IJ}_{\alpha\beta}\, \mathcal{M}_{IJ}\, ,~~\mbox{which can be split to}~~
   \{Q_\alpha, Q_\beta\}=-\frac{i}{2}(\gamma^{ab})_{\alpha\beta}\, \mathcal{M}_{ab} +  i\gamma^a\, \mathcal{P}_a\, .
\end{equation}
These conventions were directly borrowed from \cite{Zumino:1977av} and \cite{Nicolai:1984hb}, which are a little bit different than used in published paper \cite{Durka:2009pf}. One can easily check that the super--Jacobi identities are fulfilled for 
$$[ b_1,\{f_2,f_3\}]+ \{f_2,[f_3,b_1]\}-\{f_3,[b_1,f_2]\}=0\,,$$ 
as well as for the rest of possible combinations of fermionic (f) and bosonic (b) generators with pluses between all brackets. The $OSp(1,4)$ algebra can be contracted to the super-Poincar\'e algebra. To see this rescale $\mathcal{P}_a \rightarrow \ell\, \mathcal{P}_a$ with $Q \rightarrow \sqrt \ell\, Q$ and then let $\ell\rightarrow\infty$.

We now gauge the super-algebra by associating with each generator of the $OSp(1,4)$ algebra a gauge field and a gauge transformation parameter. Since the canonical dimension of the gauge field is $[-1]$ and because the dimension of gravitino $\psi_\mu$ is $[-3/2]$, as in the bosonic case, we introduce the constant $\kappa$ of the dimension $[1/2]$ fulfilling
\begin{equation}\label{c3-11}
    \kappa^2=\frac{4\pi G}{\ell}
\end{equation}  
to make the dimensions right. The gauge field is therefore
\begin{equation}\label{c3-12}
\mathbb{A}_{\mu}=\frac{1}{2}\omega^{ab}_{\mu}\mathcal{M}_{ab}+\frac{1}{\ell}e^a_{\mu}\mathcal{P}_a+\kappa\bar\psi^\alpha_{\mu} Q_\alpha\,.
\end{equation}
The curvature (\ref{c3-5}) splits into bosonic and fermionic parts
\begin{equation}\label{c3-13}
    \mathbb{F}_{\mu\nu}=\frac12\, F^{(s)}_{\mu\nu}{}^{IJ}\, \mathcal{M}_{IJ} +  \bar{\mathcal F}_{\mu\nu}^\alpha Q_\alpha=\frac12\, F^{(s)}_{\mu\nu}{}^{ab}\, \mathcal{M}_{ab}+ F^{(s)}_{\mu\nu}{}^{a}\, \mathcal{P}_{a}+ \bar{\mathcal F}_{\mu\nu}^\alpha Q_\alpha\,.
\end{equation}
First ones explicitly read as
\begin{align}\label{c3-14}
F^{(s)}_{\mu\nu}{}^{ab} &=F_{\mu\nu}^{ab}-\kappa^2\,
\bar\psi_\mu\gamma^{ab}\psi_\nu\, ,\\
    F^{(s)}_{\mu\nu}{}^a &=F_{\mu\nu}^{a}+\kappa^2\, \bar\psi_\mu\gamma^{a}\psi_\nu\,,\label{c3-14b}
\end{align}
with the AdS curvature $F_{\mu\nu}^{ab} =R^{ab}_{\mu\nu}+\frac{1}{\ell^2}( e^a_\mu e^b_\nu- e^a_\nu e^b_\mu)$ and the torsion $\ell F_{\mu\nu}^{a}=D^\omega_\mu e^a_\nu -D^\omega_\nu
e^a_\mu\,$.\\
Fermionic curvature $\mathcal F_{\mu\nu}$, with the help of covariant derivative defined to be
\begin{align}
     \mathcal D_{\mu}\psi_\nu &=
\partial_{\mu}\psi_\nu+\frac{1}{4}\omega^{ab}_\mu\,\gamma_{ab}\,\psi_\nu+\frac{1}{2\ell}
e^{a}_{\mu}\,\gamma_{a}\,\psi_\nu=\mathcal
D^\omega_{\mu}\psi_\nu+\frac{1}{2\ell}
e^{a}_{\mu}\,\gamma_{a}\,\psi_\nu \label{c3-15}\\
   \mathcal D_{\mu}\bar{\psi}_\nu&= \partial_{\mu}\bar{\psi}_\nu-\frac{1}{4}\omega^{ab}_\mu\,\bar{\psi}_\nu\,\gamma_{ab}-\frac{1}{2\ell} e^{a}_{\mu}\,\bar{\psi}_\nu\,\gamma_{a}=\mathcal
D^\omega_{\mu}\bar\psi_\nu-\frac{1}{2\ell}
e^{a}_{\mu}\,\gamma_{a}\,\bar\psi_\nu\,,\label{c3-15b}
\end{align}
can be given in a compact form as
\begin{align}\label{c3-16}
{\mathcal F}_{\mu\nu} &=\kappa\left( D_\mu\psi_\nu - D_\nu\psi_\mu\right)=\kappa \left(\mathcal D^\omega_\mu\psi_\nu
-\mathcal D^\omega_\nu\psi_\mu +\frac{1}{2\ell}\left(e_\mu^a\,
\gamma_a\psi_\nu -e_\nu^a\, \gamma_a\psi_\mu\right)\right)\,.
\end{align}

\section{Supergravity transformations and the Lagrangian}

To construct the super-BF action we also have to define another two form field $\mathbb{B}_{\mu\nu}$, which gauge-transforms in exactly the same way the curvature $\mathbb{F}_{\mu\nu}$ does. Therefore, before turning to the construction of the action we need an explicit form of gauge transformations of the components of connection and
curvature. The infinitesimal gauge transformations of the gauge field are defined in terms of the covariant derivative
\begin{equation}\label{c3-17}
\delta_\Theta \mathbb{A}_{\mu}=\partial_\mu \Theta
-i[\mathbb{A}_{\mu}, \Theta]\equiv D^{\mathbb{A}}_\mu \Theta\, ,
\end{equation}
where the gauge parameter $\Theta$ decomposes into parameters of
local Lorentz, translation and supercharge symmetries
\begin{equation}\label{c3-18}
    \Theta=\frac{1}{2}\lambda^{ab}\mathcal{M}_{ab}+\xi^a \mathcal{P}_a+\bar\epsilon^\alpha Q_\alpha\,.
\end{equation}
Using this formula one can immediately derive the supersymmetry transformations
\begin{equation}
    \delta_\epsilon e_{\mu}^{a} = -\ell\kappa\,\bar\epsilon\, \gamma^{a}\,\psi\,,\qquad
\delta_\epsilon \omega_{\mu}^{ab} = \kappa\,\bar\epsilon\, \gamma^{ab}\,\psi\mu\, ,\qquad
\delta_\epsilon \bar{\psi}_{\mu} = \frac{1}{\kappa}(\mathcal D^{\omega}_\mu\bar\epsilon-\frac{1}{2l}e^a_\mu\bar\epsilon\gamma_a)    \, ,
\end{equation}
where
\begin{equation}
    D^\omega_\mu\bar\epsilon=\partial_\mu\bar\epsilon-\frac14\omega_\mu{}^{ab}\, \bar\epsilon\gamma_{ab}\,.
\end{equation}
The supersymmetry transformations of the curvatures can be easily obtained from
\begin{equation}\label{c3-19}
\delta \mathbb{F}_{\mu\nu} = D_\mu \delta\mathbb{A}_{\nu} -D_\nu \delta\mathbb{A}_{\mu} = [D_\mu, D_\nu]\Theta=i[\Theta,\mathbb{F}_{\mu\nu}]\,, 
\end{equation}
therefore we have
\begin{equation}\label{c3-20}
 \delta_\epsilon F_{}^{(s)a4} = -\bar\epsilon \gamma^{a} {\mathcal F}_{}\,\qquad
\delta_\epsilon F_{}^{(s)ab} = \bar\epsilon \gamma^{ab} {\mathcal F}_{}\,,
\end{equation}
\begin{equation}\label{c3-21}
\delta_\epsilon \bar{\mathcal F}_{} = -\frac{1}{4}\bar\epsilon \gamma^{ab}F^{(s)}_{ab}-\frac{1}{2}\bar\epsilon\gamma_a F^{(s)a}\,.
\end{equation}

With these at hands we can now address the problem of constructing the desired supersymmetric extension of the Lagrangian (\ref{ch3-1}), and (\ref{ch3-7}). 

Let us first consider the topological theory, whose bosonic part is given by first two terms in (\ref{ch3-1}). We introduce the fermionic partner of the bosonic field $B_{\mu\nu}^{(s)IJ}$, which we denote as ${\mathcal B}_{\mu\nu}$ so that the Lagrangian reads
$$
16\pi \mathcal{L}^{(sugra-topological)} =  16\pi \left(\mathcal{L}^{(sugra-topological,b)}-4 \mathcal{L}^{(sugra-topopological,f)}\right)
$$
\begin{equation}\label{c3-22}
=\epsilon^{\mu\nu\rho\sigma} \left( B_{\mu\nu}^{(s)IJ}\, F^{(s)}_{\rho\sigma}{}_{IJ} - \frac\beta2\, B_{\mu\nu}^{(s)IJ}\, B_{\rho\sigma\,IJ}^{(s)}\right)
   -4\,\epsilon^{\mu\nu\rho\sigma} \left( \bar{\mathcal B}_{\mu\nu}{\mathcal F}_{\rho\sigma}- \frac\beta2\,\bar{\mathcal B}_{\mu\nu}{\mathcal B}_{\rho\sigma}\right)\, .
\end{equation}
This Lagrangian is invariant under local supersymmetry for the components of the field ${\mathbb B}=(B^{(s)}, {\mathcal B})$ transforming as follows
\begin{equation}\label{c3-23}
 \delta_\epsilon B_{}^{(s)a4} = -\bar\epsilon \gamma^{a} {\mathcal B}_{}\,\qquad
\delta_\epsilon B_{}^{(s)ab} = \bar\epsilon \gamma^{ab} {\mathcal B}_{}\,,\qquad
\delta_\epsilon \bar{\mathcal B}_{} = -\frac{1}{4}\bar\epsilon \gamma^{ab}B^{(s)}_{ab}-\frac{1}{2}\bar\epsilon\gamma_a B^{(s)a}\,.
\end{equation}
The gauge breaking term is invariant only under the action of the $SO(1,3)$ Lorentz subalgebra of the original gauge algebra $SO(2,3)$. Its supersymmetric extension is expected to be
\begin{equation}\label{c3-24}
16\pi \mathcal{L}^{sugra-gb} =   -\frac\alpha4\, \epsilon^{\mu\nu\rho\sigma} \left( \epsilon_{abcd} \, B_{\mu\nu}{}^{(s)ab}\, B_{\rho\sigma}^{(s)cd} -8 \, \bar{\mathcal B}_{\mu\nu}\gamma^5{\mathcal B}_{\rho\sigma}\right)\, .
\end{equation}
This term, however, is not invariant under the supersymmetry transformations given in (\ref{c3-23}), since under the latter the second term in (\ref{c3-24}) gets the contribution of the form
\begin{equation}\label{c3-25}
2\alpha\epsilon^{\mu\nu\rho\sigma} B_{\mu\nu}^{(s)a}\bar\epsilon\gamma_a\gamma^5{\mathcal B}_{\rho\sigma}
\end{equation}
that does not cancel with the transformation of the first term. As we discuss below this breaking of supersymmetry, related to the breaking of the anti-de Sitter group down to its Lorentz subgroup, does not prevent the final action from having the local supersymmetry invariance.

\section{$\mathcal{N}=1$ supergravity }

Let us now check explicitly that our procedure indeed provides the Lagrangian of ${\mathcal N}=1$ supergravity. Our starting point will be the sum of the terms (\ref{c3-22}) and (\ref{c3-24}). Field equations for the bosonic $B_{\mu\nu}{}^{IJ}$ result in the expressions analogous to (\ref{ch3-2}) and (\ref{ch3-4}):
\begin{equation}\label{c3-26}
   B_{\rho\sigma}^{(s)a}=\frac{1}{\beta}\left(F_{\rho\sigma}^{a}+\kappa^2\bar{\psi}_\rho\,\gamma^{a}\,\psi_\sigma\right)\, ,
\end{equation}
    \begin{equation}\label{c3-27}
    B_{\rho\sigma}^{(s)ab}=\frac{\beta}{\alpha^2+\beta^2} \left(F_{\rho\sigma}^{ab}-\kappa^2\bar{\psi}_\rho\,\gamma^{ab}\,\psi_\sigma\right)
    -\frac{\alpha}{2(\alpha^2+\beta^2)}\, \left(F_{\rho\sigma}^ {cd}-\kappa^2\bar{\psi}_\rho\,\gamma^{cd}\,\psi_\sigma\right)\epsilon^{ab}{}_{cd}\, ,
\end{equation}
while from their fermionic counterpart we obtain
\begin{align}\label{c3-28}
\mathcal{B}&=\frac{1}{\alpha^2+\beta^2}\left(\beta\bbone
-\alpha\,\gamma^5\, \right)\mathcal{F}\,. 
\end{align}
These substituted back to the action immediately result in the Lagrangian being sum of the fermionic and the bosonic parts
\begin{align}\label{c3-29}
16\pi \mathcal
L^{f}&=\epsilon^{\mu\nu\rho\sigma}\frac{\alpha}{(\alpha^2+\beta^2)}\,\bar{\mathcal{F}}_{\mu\nu}\left(
\frac{{\beta}\bbone
-{\alpha}\gamma^5}{2{\alpha}}\right)\,\mathcal{F}_{\rho\sigma}
\end{align}
\begin{align}\label{c3-30}
16\pi \mathcal
L^{b}&=\epsilon^{\mu\nu\rho\sigma}\left(\frac{1}{\beta}
F^{(s)a4}{}_{\mu\nu}  F_{a4}^{(s)}{}_{\rho\sigma}+\frac{1}{4}
M^{abcd} F_{ab}^{(s)}{}_{\mu\nu}
F_{cd}^{(s)}{}_{\rho\sigma}\right)
\end{align}
Up to total derivatives the term
\begin{align}\label{c3-31}
&    \bar{\mathcal{F}}_{\mu\nu}\left( \frac{\bbone{\beta} -\gamma^5{\alpha}}{2{\alpha}}\right)\,\mathcal{F}_{\rho\sigma}\,\epsilon^{\mu\nu\rho\sigma}=4\frac{\kappa^2}{2}
\mathcal (D^{}_\mu \bar\psi_\nu)\,(\gamma\bbone -\gamma^5)\mathcal (D^{}_\rho \psi_\sigma)\,\epsilon^{\mu\nu\rho\sigma}
\end{align}
after some straightforward but tedious calculations can be rewritten as
\begin{align}
\bar{\mathcal{F}}_{\mu\nu}\left( \frac{\bbone{\beta} -\gamma^5{\alpha}}{2{\alpha}}\right)\,\mathcal{F}_{\rho\sigma}\,\epsilon^{\mu\nu\rho\sigma}    &=\frac{\kappa^2}{4}\bar\psi_\mu\,(\gamma\bbone -\gamma^5)\, \left(\gamma_{ab} \,F^{ab}_{\nu\rho}+\gamma_a\frac{2}{\ell}T_{\nu\rho}^a\right)\,\psi_\sigma\,\epsilon^{\mu\nu\rho\sigma}\nonumber\\
    &+\kappa^2\,\bar\psi_\mu\,\left(\frac{1}{\ell^2}\gamma^5\gamma_{ab}\,e^a_\nu\, e^b_\rho+\frac{2}{\ell}\gamma^5\gamma_{a}\,e^a_\nu\,\mathcal D^\omega_\rho\right)\psi_\sigma\,\epsilon^{\mu\nu\rho\sigma}\,.\label{c3-32}
   \end{align}
It brings the total Lagrangian to the following form
\begin{align}
  16\pi \mathcal L&=-\left(\frac{\kappa^2}{G}\,\bar\psi_\mu\,\gamma^5\gamma_{ab}\,e^a_\nu\, e^b_\rho+ \frac{2\kappa^2\ell}{G}\,\bar\psi_\mu\,\gamma^5\gamma_{a}\,e^a_\nu\,\mathcal D^\omega_\rho\psi_\sigma\right)\,\epsilon^{\mu\nu\rho\sigma}\nonumber\\
&-\bar\psi_\mu\,\left(\frac{1}{4\beta} \frac{2\kappa^2}{\ell}\gamma_{a}\,T_{\nu\rho}^a+\frac{2\kappa^2\ell}{4G}\,(\gamma\bbone -\gamma^5)\, \gamma_a\,T_{\nu\rho}^a\right)\,\psi_\sigma\,\epsilon^{\mu\nu\rho\sigma}\nonumber\\
&-\frac{1}{4\beta} \left(\frac{1}{\ell^2}T_{\mu\nu}^{a}\,T_{\rho\sigma\,a}+\kappa^4 \,\bar\psi_\mu\gamma^{a}\psi_\nu \, \bar\psi_\rho\gamma_{a}\psi_\sigma\right)\,\epsilon^{\mu\nu\rho\sigma}\nonumber\\
&+\frac{1}{16} M_{abcd} \left(F_{\mu\nu}^{ab}\,F_{\rho\sigma}^{cd}+\kappa^4\, \bar\psi_\mu\gamma^{ab}\psi_\nu\, \bar\psi_\rho\gamma^{cd}\psi_\sigma\right)\,\epsilon^{\mu\nu\rho\sigma}\label{c3-33}\\
&+\mbox{\em total derivative}\,.\nonumber
\end{align}
Notice, that by exploiting identity $\gamma_{ab}\gamma^5=\frac{1}{2}\epsilon_{abcd}\gamma^{cd}$ we already made the cancellations between $\mathcal{L}^f $ and $\mathcal{L}^b$ of the terms with curvature $F^{ab}$ with quadratic spinors. By making use of the Fierz identities with $
 \epsilon^{\mu\nu\rho\sigma}\, \bar\psi_\mu\, \Gamma\, \psi_\nu\, \Gamma\, \psi_\rho =0\, ,
 $
where $\Gamma$ is an arbitrary combination of $\gamma$ matrices, and
$$
\epsilon^{\mu\nu\rho\sigma}\,\bar\psi_\mu\, \Gamma^A\, \psi_\nu=0, \quad \mbox{for}\quad\Gamma^A=(1,\gamma^5,\gamma^5\gamma^a)\, ,
$$
one can check that four-fermion terms vanishes identically as well, along with some simplifications in the second line. The Lagrangian reduces therefore to the final form
\begin{align}
&   16\pi \mathcal L=\left(
\frac{1}{16} M_{abcd} \,F_{\mu\nu}^{ab}\,F_{\rho\sigma}^{cd}-\frac{1}{4\beta\ell^2} T_{\mu\nu}^{a}\,T_{\rho\sigma\,a}\right)\,\epsilon^{\mu\nu\rho\sigma}\nonumber\\
&-\left(\frac{\kappa^2}{G}\,\bar\psi_\mu\,\gamma^5\gamma_{ab}\,e^a_\nu\, e^b_\rho+\frac{2\kappa^2\ell}{G}\,\bar\psi_\mu\,\gamma^5\gamma_{a}\,e^a_\nu\,\mathcal D^\omega_\rho\psi_\sigma\right)\,\epsilon^{\mu\nu\rho\sigma}\nonumber\\
&+\frac{\kappa^2\ell}{2\gamma G}\,\bar\psi_\mu
\gamma_{a}\,\psi_\nu\,T_{\rho\sigma}^a \,\epsilon^{\mu\nu\rho\sigma}
+\mbox{\em total derivative}\,.\label{c3-34}
\end{align}
We can find the Lagrangian that can be decomposed into three types of terms. The expansion of the curvature $F_{\mu\nu}{}^{ab}$ obviously leads to the Einstein--Cartan Lagrangian (\ref{Einstein-Cartan}) with cosmological constant, to which we add the second line to form standard supergravity Lagrangian. We fix $\kappa^2= 4\pi G/\ell$ to make the coefficient of the gravitino kinetic term equal $1/2$, and we end with exactly what can be found in \cite{Townsend:1977qa}:
\begin{align}
 \mathcal{L}^{sugra}&=\frac{1}{64\pi G} \left(R_{\mu\nu}^{ab}\, e_\rho{}^c\, e_\sigma{}^d +\frac{1}{\ell^2}\, e_\mu{}^{a}\,e_\nu{}^{a}\, e_\rho{}^c\, e_\sigma{}^d \right) \,\epsilon_{abcd}\,\epsilon^{\mu\nu\rho\sigma}\nonumber\\
 &+ \left(\frac{1}{2} \, \bar{\psi}_\mu\,\gamma_5\,\gamma_{a}\,e^{a}_{\nu}D^\omega_{\rho}\psi_\sigma\, +\frac{1}{4\ell} \, \bar{\psi}_\mu\,\gamma_5\,\gamma_{ab}\,e^{a}_{\nu}e^{b}_{\rho}\,\psi_\sigma \right) \epsilon^{\mu\nu\rho\sigma}\,.\label{c3-35}
\end{align}
The second class of terms contains the Holst term and it's supersymmetric counterpart, combined into additional Lagrangian
\begin{equation}\label{c3-36}
 \mathcal{L}^{add}= \frac{1}{\gamma}\,\left(\frac{2}{64\pi G} \, R_{\mu\nu}{}^{ab}\, e_\rho{}_a\,e_\sigma{}_b + \frac{1}{4}\,\bar{\psi}_\mu\, \gamma_{a}\, \psi_\nu \, D^\omega_\rho e_\sigma{}^a\right)\, \epsilon^{\mu\nu\rho\sigma}\, .
\end{equation}
The remaining terms (Euler, Pontryagin, and Nieh-Yan) can be added to the total derivatives coming from fermionic part. Thus, we find that resulting boundary term is being expressed by the combination of super Chern-Simons terms (for the connections $A^{IJ}$, antiselfdual, and selfdual ${}^\pm\omega^{ab}$) with some additional fermionic current
\begin{align}
\frac{32\pi G}{\ell^2}\mathcal{L}^{boundary}&= \partial_\mu\Big[ \Big(\frac{1}{\gamma}+i\Big)\mathcal{SC}^\mu (^+\omega)+ \Big(\frac{1}{\gamma}-i\Big)\mathcal{SC}^\mu (^-\omega)-\big(\frac{1}{\gamma} +\gamma\big)\mathcal{SC}^\mu(A)\Big]\nonumber\\
&+4\kappa^2\partial_\mu \Big[  \,\bar{\psi}_\nu\frac{1}{2}(\frac{1}{\gamma}\mathbb{I}-\gamma^5)\frac{1}{2}A_\rho^{IJ}m_{IJ} \psi_\sigma\Big)\epsilon^{\mu\nu\rho\sigma}\Big]\,.\label{c3-37}
\end{align}
In above we used the definition of the super Chern-Simons terms
\begin{align}
 \mathcal{SC}^\mu(A)&=\Big(A_{\nu IJ}\partial_{\rho} A_{\sigma}^{IJ}+\frac{2}{3} A_{\nu IJ}A_{\rho}^I\,_K A_{\sigma}^{KJ}\Big)\epsilon^{\mu\nu\rho\sigma}+4\kappa^2\left(\bar{\psi}_\nu\,D^A_\rho\psi_\sigma\right)\,\epsilon^{\mu\nu\rho\sigma}\nonumber\\
&=\mathcal{C}^\mu(A)+4\kappa^2\left(\bar{\psi}_\nu\,D^A_\rho\psi_\sigma\right)\,\epsilon^{\mu\nu\rho\sigma}
\label{c3-38}\\
 \mathcal{SC}^\mu({}^\pm\omega)&= \mathcal{C}^\mu({}^\pm\omega)+4\kappa^2\left({}^\pm\bar{\psi}_\nu\,D^{{}^\pm\omega}_\rho{}^\pm\psi_\sigma\right)\,\epsilon^{\mu\nu\rho\sigma}
\label{c3-39}
\end{align}
Although such a form of the boundary (established with M. Szczachor) looks quite interesting, it still remains unclear. We will leave this subject to some separate paper, and now return to the Immirzi parameter being main objective of this chapter.

\section{Influence of the Immirzi parameter on supergravity}

After deriving the form of the Lagrangian we would like to check if the action obtained from (\ref{c3-35}) and (\ref{c3-36}) is indeed invariant under supersymmetry. To do that we make use of the $1.5$ formalism (see \cite{VanNieuwenhuizen:1981ae} and references therein), which combines the virtues of the first ($\omega$ is an independent field) and second ($\omega=\omega(e,\psi)$) order formalisms. The idea is as follows. Our action $I$, being the integral of the Lagrangian can be thought of as a functional $I(e;\psi; \omega(e,\psi))$ and its variation is
\begin{equation}
    \delta I = \delta e\, \left.\frac{\delta I}{\delta e}\right|_{\psi, \omega(e,\psi)} + \delta \psi\, \left.\frac{\delta I}{\delta \psi}\right|_{e, \omega(e,\psi)}
+\left.\frac{\delta I}{\delta \omega}\right|_{e,\psi}\left(\delta e\,\frac{\delta \omega(e,\psi)}{\delta e}+ \delta \psi\,\frac{\delta \omega(e,\psi)}{\delta \psi}\right) \, .\label{c3-40}\end{equation}
But if $\omega$ satisfies its own field equations the last term in (\ref{c3-40}) vanishes identically, because $\delta I/\delta\omega=0$ for $\omega$ satisfying its own field equation. In other words we need to vary only the gravitino and tetrad fields, taking into account, where necessary, the conditions coming from the Lorentz connection field equations.

The first step in checking the supersymmetry is therefore to analyze form of the $\omega$ field equations
\begin{equation}\label{c3-41}
\left(\frac{1}{\ell}T_{\mu\nu}^{a}+\kappa^2\, \bar\psi_\mu\gamma^{a}\psi_\nu\right)\,e^b_\rho \left( \frac{1}{\gamma}\delta_{abcd}+\epsilon_{abcd}\right)\epsilon^{\mu\nu\rho\sigma}\,\delta\omega^{cd}_\sigma=0
\end{equation}
where the first bracket is nothing else than "supertorsion"' $F^{(s)a}$ defined in (\ref{c3-14}). It follows from (\ref{c3-41}), that for $\displaystyle \gamma+\frac{1}{\gamma}\neq 0$ (reduced down to $\gamma^2\neq-1$) the supertorsion vanishes. 
To see this just contract this field equation by $\epsilon_{abcd}$, and use the fact that tetrad is invertible. Thus one finds
\begin{equation}\label{c3-42}
  T_{\mu\nu}^{a}=-4\pi G \, \bar\psi_\mu\gamma^a\psi_\nu\, ,
\end{equation}
which allows to extract the connection $\omega(e,\psi)$ from the condition
\begin{equation}\label{c3-43}
\partial_\mu e_\nu^{a}   -\partial_\nu e_\mu^{a} + \omega_\mu{}^a{}_b\, e_\nu^b - \omega_\nu{}^a{}_b\, e_\mu^b+4\pi G \, \bar\psi_\mu\gamma^a\psi_\nu\,=0 ,
   \end{equation}
as it was shown in \cite{VanNieuwenhuizen:1981ae}. It's crucial because by vanishing of the super-torsion above we are able to fulfill supersymmetry of a starting action.\\ \indent 
Since the Lorentz connection field equations are the same as in the standard case of ${\cal N}=1$ supergravity, it is just a matter of repeating the steps described in \cite{Townsend:1977qa}, \cite{Ferrara:1976kg}, and \cite{VanNieuwenhuizen:1981ae} to see that (\ref{c3-35}) is indeed supersymmetric (up to total derivative term). What remains therefore is to check if $\mathcal{L}^{add}$ (\ref{c3-36}) is supersymmetric as well. But this is also quite straightforward. In fact one can prove a much stronger result, namely that {\em if supertorsion is zero, for arbitrary $\delta e$, $\delta \psi$ the variation of $\mathcal{L}^{add}$ vanishes}. This not only proves the supersymmetry invariance but also shows that the Lagrangian $\mathcal{L}^{add}$ does not contribute to the field equations.\\\indent
Consider the variation of the gravitino first to find from $\mathcal{L}^{add}$ the term
\begin{equation}\label{c3-44}
  \frac{1}{4\gamma}\,\epsilon^{\mu\nu\rho\sigma}\,  \delta\bar{\psi}_\mu\, \gamma_{a}\, \psi_\nu \, D^\omega_\rho e_\sigma{}^a\, .
\end{equation}
But since $D^\omega{}_{[\rho} e_{\sigma]}{}^a\sim \bar\psi_\rho\gamma^a\psi_\sigma$ (due to vanishing of supertorsion (\ref{c3-41})), and by the identity
\begin{equation}
 \epsilon^{\mu\nu\rho\sigma}\, \gamma_a\psi_\nu\,\bar\psi_\rho\gamma^a\psi_\sigma=0   
\end{equation}
we see that this expression vanishes. Thus it remains to check the variation of tetrad
\begin{equation}\label{c3-45}
\frac{1}{\gamma}\,\left(\frac{2}{8\pi G} \, R_{\mu\nu}{}^{ab}\, e_\rho{}_a\,\delta e_\sigma{}_b + \bar{\psi}_\mu\, \gamma_{a}\, \psi_\nu \, D^\omega_\rho \delta e_\sigma{}^a\right)\, \epsilon^{\mu\nu\rho\sigma}\, .
\end{equation}
Making use of the second Bianchi identity
\begin{center}
    $\displaystyle
    \epsilon^{\mu\nu\rho\sigma}\, R_{\mu\nu}{}^{ab}\, e_\rho{}_a=-2\epsilon^{\mu\nu\rho\sigma}\, D^\omega_\mu\, D^\omega_\nu e_\rho{}^b\,,
$
\end{center}
forces (\ref{c3-45}) to be (up to the total derivative) rewritten as
\begin{equation}\label{c3-46}
    \frac{1}{\gamma}\,\left(\frac{1}{4\pi G} \,\, T_{\nu\rho}^b\,D^\omega_\mu\delta e_\sigma{}_b + \bar{\psi}_\mu\, \gamma_{a}\, \psi_\nu \, D^\omega_\rho \delta e_\sigma{}^a\right)\, \epsilon^{\mu\nu\rho\sigma}\, .
\end{equation}
which, with the help of vanishing of supertorsion can be easily seen to vanish. This not only completes the proof of supersymmetry, but it also shows that the terms (\ref{c3-36}) do not contribute to field equations. It finds the support in the results of \cite{Kaul:2007gz}. However it should be stressed that although invisible classically the term $\mathcal{L}^{add}$ might be relevant in quantum theory like the QCD theta term.

It is worth noticing that the proof of supersymmetry of the final supergravity Lagrangian $\mathcal{L}^{sugra} + \mathcal{L}^{add}$ makes it possible to resolve the puzzle that we encountered earlier. Namely if we make use of the fact that $B_{\mu\nu}^{a}$ equals supertorsion, and that the latter vanishes, we see that the expression (\ref{c3-26}) is zero. This is why the apparent lack of supersymmetry of the considered theory does not prevent the final one from being supersymmetric.

This ends present chapter. Theory of gravity seen as a deformation of $BF$ theory extends notion of ${\cal N}=1$ supergravity to the case of the presence of Immirzi parameter, but shows no influence from it in the final outcome. Let's now try to challenge the similar problem in the context of black hole thermodynamics.


%
\chapter{Black hole thermodynamics and the Immirzi parameter}
This chapter is devoted to establish comparison between results coming from black hole thermodynamics of LQG, and an approach corresponding to the gravitational Noether charges in the first order gravity. As we will see, the framework of $BF$ theory offers interesting setting serving this purpose. This subject contributes to the motivation of a taken model, and brings some interesting results.

\section{Black hole thermodynamics}

The discovery of the laws of black hole dynamics \cite{Bardeen:1973gs} has led to uncovering a remarkable analogy between gravity and thermodynamics. This is especially clear in the case of the first law
\begin{equation}\label{c5-1}
 dM = \frac{\kappa}{8\pi G}\,dA+\Omega dJ \mathrm{~(black~hole~dynamics)}
\end{equation}
\begin{equation}
     dE=TdS+dW   \mathrm{~(thermodynamics)}\,, 
\end{equation}
where in the first line we have black hole mass $M$, angular momentum $J$, angular velocity $\Omega$, area of the event horizon $A$, and surface gravity $\kappa$.

The seminal works of Bekenstein \cite{Bekenstein:1973ur, Bekenstein:1974ax} and Hawking \cite{Hawking:1974sw}, relating the area of the event horizon with the entropy
\begin{equation}\label{c5-2}
   Entropy= \frac{Area}{4l_p^2}, \qquad\qquad \mathrm{where}\qquad  l_p=\sqrt{G\hbar/c^3}\,,
\end{equation}
and the surface gravity of a black hole with its temperature
\begin{equation}
    Temperature=\hbar c\,\frac{\kappa}{2\pi}\,,
\end{equation} 
have strongly suggested that this analogy might be in fact an identity. Yet, it still lacks better and deeper understanding \cite{Jacobson:2007uj, Kolekar:2010dm, Verlinde:2010hp}.

A major step in this direction has been done by Robert Wald, who showed that gravitational quantities from (\ref{c5-1}) could be obtained as the Noether charges \cite{Wald:1993nt, Iyer:1995kg}. 

The entropy can be calculated in various approaches to quantum gravity. In particular, within the loop quantum gravity approach, it turns out that key ingredient, the Immirzi parameter \cite{Immirzi:1996di, Holst:1995pc}, is explicitly present in the black hole entropy formula
\begin{equation}\label{c5-3}
    S_{LQG}=\frac{\gamma_M}{\gamma}\;\frac{Area}{4G}\,,
\end{equation}
where $\gamma_M$ is a numerical parameter valued between
$0.2$ and $0.3$. This result, coming from microscopic description and counting microstates, agrees with Bekenstein's entropy only when $\gamma$ is fixed to get rid off a numerical prefactor (\cite{Rovelli:1996dv}, \cite{Domagala:2004jt}, \cite{Meissner:2004ju} for recent review see \cite{Agullo:2010zz}). Such transition is poorly understood and must be further explored (see however \cite{Ghosh:2011fc}). 

We will focus on Wald's procedure applied to first order gravity in the setting, where the Holst modification and the Immirzi parameter are present in the action we start with. The deformed $SO(2,3)$ $BF$ theory allows us to check if the Noether approach is able to reproduce the result obtained in the LQG framework. After deriving generalized formulas for the gravitational charges \cite{Durka:2011yv} we will investigate a few AdS spacetimes  \cite{Durka:2011zf} (Schwarzschild, topological black holes, Kerr, Taub--NUT), and discuss the outcome in the context of the Immirzi parameter.

\section{Wald's approach and gravitational Noether charges}

Emmy Noether's theorem concerning differentiable symmetries of the action of a physical system and the resulting conservation laws, holds well deserved place in theoretical physics.

If one considers a variation of the action, then, except the field equations $(f.e.)$ multiplied by variated field, a boundary term $\Theta$ arises
$$\delta \mathcal L(\varphi,\partial \varphi)=(f.e.)\cdot \delta \varphi +d\Theta \,. $$
For any diffeomorphism $\delta_\xi\varphi=L_\xi \varphi$ being generated by a smooth vector field $\xi^\mu$, we can derive a conserved Noether current
$$
    J[\xi] = \Theta[\varphi, L_\xi \varphi] - I_\xi \mathcal L\,,
$$
where $L_\xi$ denotes the Lie derivative in the direction $\xi$ and the contraction operator $I_\xi$ acting on a p-form is given by $I_\xi \alpha_p=\frac{1}{(p-1)!}\xi^\mu \alpha_{\mu\nu^1...\nu^{p-1}}dx^{\nu^1}\wedge...\wedge dx^{\nu^{p-1}}$.

Noether current is closed on shell, which allows us to write it in the terms of the Noether charge 2-form $Q$ by the relation $J = dQ $.

In Wald's approach one applies this construction to the Einstein-Hilbert Lagrangian and its diffeomorphism symmetry \cite{Wald:1993nt, Iyer:1995kg}. The generators, being Killing vectors associated with the time translations ($\xi_t=\partial_t$) and spacial rotations ($\xi_\varphi=\partial_\varphi$), by the integration at the infinity, produce $Q[\xi_t]_{\infty}$ and $Q[\xi_\varphi]_{\infty}$, which are mass and angular momentum, respectively. Charge calculated for the horizon generator $$Q[\xi_t+\Omega\xi_\varphi]_{H}=\frac{\kappa}{2\pi}\,Entropy$$ ensures the product of the horizon temperature, and the entropy.

\section{First order formulation and topological regularization of charges}

The outcome of this formalism agrees with other methods in different frameworks, however such a procedure applied directly to the tetrad formulation of gravity (\ref{Einstein-Cartan})
leads to serious problems. Noether charge evaluated for AdS--Schwarzschild metric gives wrong factor before $M$ and also requires background subtraction to get rid off cosmological divergence
\begin{equation}\label{c5-4}
    Mass=Q[\xi_t]_\infty=\frac{1}{2}\,M+\lim_{r\rightarrow\infty}\frac{r^3}{2G\ell^2}.
\end{equation}
To deal with this apparent problem it was suggested by Aros, Contreras, Olea, Troncoso and Zanelli \cite{Aros:1999id, Aros:1999kt} to use the Euler term as the boundary term
\begin{equation}\label{c5-5}
     32\pi G\, S=\int R^{ab}\wedge e^{c}\wedge e^{d}\,\epsilon_{abcd}+\frac{1}{\,2\ell^2}\int  e^{a}\wedge e^{b}\wedge e^{c}\wedge e^{d}\, \epsilon_{abcd}+\rho\int R^{ab}\wedge R^{cd} \,\epsilon_{abcd}\,,
\end{equation}
so for arbitrary weight $\rho$ the formula (\ref{c5-4}) changes into
\begin{equation}\label{c5-6}
Mass=Q[\partial_t]_\infty=\frac{M}{2}\left(1+\frac{2}{\ell^2}\,\rho\right)+\lim_{r\rightarrow\infty}\frac{r^3}{2G\ell^2}\left(1-\frac{2}{\ell^2}\,\rho\right)\,.
\end{equation}
It is easy to notice, that the fixed weight $\rho=\frac{\ell^2}{2}$ cures the result, because it simultaneously removes the divergence and corrects the factor before the mass value. Remarkably this is exactly the value known from MacDowell and Mansouri prescription! Moreover, with the Euler term contributing to a boundary term, adding boundary condition of the AdS asymptotics at infinity
\begin{equation} \label{c5-7}
(R^{ab}(\omega)+ \frac1{\ell^2}\, e^a\wedge e^b)\Big|_\infty=0
\end{equation}
and fixing the connection $\delta \omega=0$ on the horizon (in order to fix a constant temperature and fulfill the zeroth law) ensures differentiability of the action 
\begin{equation}\label{c5-8}
    \delta S_{(Einstein/Cartan+\Lambda+\frac{\ell^2}{2}Euler)}=\int_{\mathcal M}(f.e.)_{a}\delta e^a+\int_{\mathcal M} (f.e.)_{ab}\,\delta \omega ^{ab}+\int_{\mathcal M}d\Theta=0\,,
\end{equation}
where $(f.e.)$ are field (Einstein and torsion) equations and boundary term 
\begin{equation}\label{c5-9}
    \Theta\Big|_{\partial\mathcal{M}}
=\epsilon_{abcd}\,\delta\omega^{ab}\wedge \left(R^{cd}(\omega)+ \frac1{\ell^2}\, e^c\wedge e^d\right)\Big|_{\partial\mathcal{M}}=0.
\end{equation}

\section{Generalized Noether charges from the deformed BF theory}

Differentiability of the action (\ref{ch3-1}) is naturally incorporated by the field equations and the boundary conditions specified above. Besides the bulk terms, and taking into account that $B^{a4}\sim T^a$ vanishes, the boundary integral reads as
\begin{equation} \label{c5-10}
  \int_{\mathcal \partial M}\delta A^{IJ}\wedge B_{IJ}\qquad \to\qquad  \int_{\mathcal \partial M}\delta\omega^{ab}\wedge B_{ab}= \int_{\mathcal \partial M}\delta\omega^{ab}\wedge M_{abcd} F^{ab}=0\,.
\end{equation} 
Despite of the different signs and form of factors related to the Immirzi parameter in (\ref{ch3-7}), we have to remember that in a derivation process the equations of motion have to be solved, which forces torsion $T^a$ to vanish, so what is left from a Nieh-Yan term adds directly to the Holst term. Then the scheme is equipped by somehow analogous combination
\begin{center}
    [Einstein/Cartan with $\Lambda$ +$\frac{\ell^2}{2}$Euler] $-\gamma$ [Holst +$\frac{\ell^2}{2}$Pontryagin],
\end{center}
which makes further extension of Wald's procedure presented earlier rather clear and straightforward. 

Now knowing that  the action (\ref{ch3-1}) is differentiable we can turn
to the discussion of the Noether charges associated with its
symmetries. In our derivation below we will follow the procedure
proposed in the papers \cite{Wald:1993nt} and \cite{Iyer:1995kg}.
Let us start with an arbitrary variation of the action (\ref{ch3-1})
$$  \begin{aligned}
\displaystyle 16\pi\,\delta S= \int \Big(&\delta B^{IJ}
\wedge (F_{IJ}-\beta B_{IJ}-\frac{\alpha}{2}B^{KL} \,\epsilon_{IJKL4})+\\
&+\delta A_{IJ} \wedge (D^A B^{IJ} ) +d( B^{IJ} \wedge \delta
A_{IJ})\Big)\,.
 \end{aligned}$$
The expressions proportional to the variations of $B^{IJ}$ and
$A^{IJ}$ in the bulk are field equations, while the last term is the total
derivative of the 3-form symplectic potential:
\begin{equation}
   \Theta=  B^{IJ} \wedge \delta A_{IJ}\,.
\end{equation}
For an arbitrary diffeomorphism generated by a smooth vector field
$\xi^\mu$, one can derive the conserved Noether current 3-form $J$
given by
\begin{equation}\label{18}
    J[\xi] = \Theta[\phi, L_\xi \phi] - I_\xi  \mathcal{L},
    \qquad  J[\xi] =B^{IJ} \wedge L_\xi A_{IJ}-I_\xi  \mathcal{L}
\end{equation}
where $\mathcal{L}$ is the Lagrangian, $L_\xi$ denotes the Lie derivative in
the direction $\xi$ and contraction $I_\xi$ (acting on a $p$-form $\alpha$) is
defined to be
$$
I_\xi \alpha_p=\frac{1}{(p-1)!}\xi^\mu\,
\alpha_{\mu\nu^1...\nu^{p-1}}dx^{\nu^1}\wedge...\wedge
dx^{\nu^{p-1}}\,.
$$
By direct calculation we find
$$  \begin{aligned}
 16\pi\,   J[\xi] &= \left(F_{IJ}-\beta B_{IJ}-\frac{\alpha}{2}B^{KL} \,\epsilon_{IJKL4}\right)\wedge I_\xi B_{IJ}\\
&+I_\xi A_{IJ}\wedge\left(D^A B^{IJ}\right)  +d\left(B^{IJ}\wedge
I_\xi A_{IJ}\right)\,.
 \end{aligned}$$
When field equations are satisfied this current is an exact
differential of a two form and thus we can write the associated
charge to be
\begin{equation}\label{19}
    Q[\xi]=\frac1{16\pi}\int_{\partial \Sigma} B^{IJ}\, I_\xi A_{IJ}
\end{equation}
which after substituting the solution for the $B$ field equations takes the form
\begin{equation}\label{20}
     Q=\frac1{16\pi}\int_{\partial \Sigma} \left(\frac{1}{2} M^{ab}{}_{cd}\, F_{ab}\, I_\xi \omega^{cd}-\frac{2}{\beta\ell^2}\,
     T_a \, I_\xi e^a\right)\,,
\end{equation}
where $\partial\Sigma$ is a spatial section of the manifold. One can  check that this expression for the Noether charge agrees with the one obtained explicitly from the first order action, as it should. 

\section{Immirzi parameter impact?}
The most general form of first order gravity coming from the connection $A^{IJ}$ and the formulation of $SO(2,3)$ $BF$ theory leads to natural generalization of the charge formula
\begin{equation}
Q[\xi]=\frac{1}{16\pi}\int_{\partial \Sigma} \frac{\delta \mathcal L}{\delta F^{IJ}}\;I_\xi A^{IJ}+(f.e.)_{IJ}\, I_\xi A^{IJ}\,.    
\end{equation}
After substituting the $B$ fields according to (\ref{ch3-2}) and (\ref{ch3-4}), and solving the field equations, which makes torsion vanish, we can write down the associated charge to be 
\begin{equation}\label{e20}
    Q[\xi]=\frac1{16\pi}\int_{\partial \Sigma} I_\xi \omega^{cd}\,\left(\frac{1}{2} M^{ab}{}_{cd}\, F_{ab}\right)\, ,
\end{equation}
The final form originally derived in \cite{Durka:2011yv}
\begin{equation}\label{e28}
   Q[\xi]=\frac{\ell^2}{32\pi G}\int_{\partial \Sigma} I_\xi \omega_{ab}
   \left(\epsilon^{ab}_{\;\;\;cd}F^{cd}_{\theta\varphi}-2\gamma F^{ab}_{\theta\varphi}\right)d\theta\, d\varphi\,.
\end{equation}
generalizes the results of \cite{Wald:1993nt,Iyer:1995kg}, and \cite{Aros:1999kt} to the case of first order gravity with the Holst modification. This formula, except choosing the AdS asymptotics, was derived without specifying any initial background. To check wherever presence of the Immirzi parameter is in fact possible we have to turn to the explicit AdS spacetimes. 

\subsection{AdS--Schwarzschild}
We begin with the standard case of a black hole in the presence of a negative cosmological constant
\begin{equation}\label{AdS-Schwarschild}
ds^2=-f(r)^2 dt^2 +f(r)^{-2} dr^2+r^2(d\theta^2+\sin^2 \theta d\varphi^2)\,,\qquad\quad f(r)^2=(1-\frac{2GM}{r}+\frac{r^2}{\ell^2})\,.    
\end{equation}
Evaluating $I_\xi \omega_{ab}$ for the timelike Killing vector immediately forces 
\begin{equation}
    Q[\partial_t]=\frac{4\ell^2}{32\pi G}\int_{\partial\Sigma} \omega^{01}_{t}\left(\epsilon_{0123}F^{23}_{\theta\varphi}-\gamma F^{}_{\theta\varphi\,01}\right)d\theta\,d\varphi\,.  
\end{equation}
For the AdS--Schwarzschild metric given in (\ref{AdS-Schwarschild}) the term $F^{01}_{\theta\varphi}$ multiplied by the Immirzi parameter is equal to zero, thus its whole modification drops out, leaving only expression for the MacDowell--Mansouri Noether charge already obtained in \cite{Aros:1999id} and \cite{Aros:1999kt}
\begin{equation}\label{AdS-Noether}
   Q[\partial_t]=\frac{4\ell^2}{32\pi G}\int_{\partial \Sigma}  \left(\frac{1}{2} \frac{\partial f(r)^2}{\partial r} \right)\left(1-f(r)^2+\frac{r^2}{\ell^2}\right)\sin \theta \,d\theta \, d\varphi\,.    
\end{equation}
Evaluating the charge associated with the timelike Killing vector at infinity returns right answer for the mass
\begin{equation}
    Q [\xi ]{}_\infty=\lim_{r\rightarrow\infty}\frac{1}{4\pi }
    \int_{\partial \Sigma_{\infty}}  \left( M+\frac{\ell^2 GM^2}{r^3}\right)
    \sin \theta \,d\theta\, d\varphi = M\,.
\end{equation}
To recover the entropy from (\ref{AdS-Noether}) we first need to introduce the surface gravity $\kappa$, which can be defined (as it was done in \cite{Aros:2005by}) by the rescaled Killing vectors $\xi^a=e^a_\mu\,\xi^\mu$ in the formula
\begin{equation}
    I_\xi \omega^{a}{}_{b}\, \xi^b=\kappa\,\xi^a
\end{equation}
being a first order analog of the standard definition ($\xi^\mu \nabla_\mu \xi^\nu=\kappa \,\xi^\nu $)
in a metric formulation. Straightforward calculations shows that for the AdS--Schwarzschild
\begin{equation}
\kappa=\omega^{01}_t \Big|_{r_H}= \left(\frac{1}{2} \frac{\partial f(r)^2}{\partial r}
\right)\Big|_{r_H} \qquad\qquad T=\frac{\kappa}{2\pi}\,,
\end{equation}
Therefore, at the horizon defined by 
$$f(r_H)^2=0,\qquad \frac{r^3_H}{\ell^2}+r_H-2G M=0$$
the charge (\ref{AdS-Noether}) becomes 
\begin{equation}
     Q[\xi_t]_H=\frac{\kappa \,\ell^2}{8\pi G}\left(1+\frac{r_H^2}{\ell^2}\right)\int_{\partial \Sigma_{H}}
\sin \theta \;d\theta\, d\varphi=\frac{\kappa}{2\pi}\frac{4\pi(r_H^2 +\ell^2)}{4G}\,,
\end{equation}
so the black hole entropy yields
\begin{equation}\label{entropy1}
    Entropy=\frac{Area}{4G}+\frac{4\pi\ell^2}{4G}\, .
\end{equation}
It differs from the standard form by a constant. This does not alter the fist law, because over there we are only interested in the change of quantities (for discussion of the second law see \cite{Liko:2007vi}). Similar result appears in the Lovelock gravities, where entropy gains the term proportional to the arbitrary factor before the Gauss-Bonnet term \cite{Iyer:1994ys} (here, at least we avoid problem of the negative entropy \cite{Clunan:2004tb}). Relation between the Euler characteristic and the entropy of extreme black holes was investigated in \cite{Liberati:1997sp}.

Resulting constant has no satisfactory interpretation. We cannot go any further than noticing, that this apparent drawback of Euler regularization has the exact value of the cosmological horizon entropy \cite{Gibbons:1977mu, Bousso:2002fq} for the pure de Sitter spacetime. The de Sitter space has the different asymptotic structure, in which instead of infinity we have the cosmological horizon. It seems justified to undertake separately this subject, as it has great importance for our observed Universe. A positive cosmological constant implies a radical change in the situation by the existence of a cosmological horizon; this means different boundary conditions, and because of the two temperatures associated with the horizons, it is necessary to provide a description of the evolving system in thermodynamic imbalance. The corresponding analysis will be presented in the future.

\subsection{Topological black holes}
Let us now focus on the topological black holes \cite{Vanzo:1997gw, Cai:1998vy,Brill:1997mf}, for which the event horizons are surfaces of nontrivial topology. The geometries of pseudo-sphere, torus, and sphere are represented by $k=-1,0,1$ in the function
\begin{equation}
    f(r)^2=(k-\frac{2GM}{r}\frac{4\pi}{\Sigma_k}+\frac{r^2}{\ell^2})\,,
\end{equation}
where $\Sigma_k$ is the unit area of the horizon hypersurface coming from the surface element
\begin{equation}
   d\Sigma_k= \left\{
\begin{array}{l l}
\displaystyle 
\sinh\theta\; d\theta \, d\phi &\mathrm{for}\qquad k=-1\\
d\theta\, d\phi &\mathrm{for}\qquad k=0\\
\sin\theta\,  d\theta \, d\phi & \mathrm{for}\qquad k=1
\end{array} \right.
\end{equation}

This generalization of geometry does not change the situation concerning the Immirzi parameter. Once again, the expression $I_\xi\omega^{ab}$ for the field $\partial_t$ forces Immirzi contribution to be of the form $F_{\varphi \theta}^{01}$, which is exactly zero. Thus, the Noether charge 
\begin{equation}
   Q[\xi]=\frac{4\ell^2}{32\pi G}\int_{\partial \Sigma} \left(\frac{1}{2}\frac{\partial f(r)^2}{\partial r}\right)
   \left(k-f(r)^2+\frac{r^2}{\ell^2}\right)\,d\Sigma_k
\end{equation}
gets the values at infinity and at the horizon
\begin{equation}
   Q[\xi_t]_\infty=\frac{M}{\Sigma_k}\int_{\partial \Sigma_\infty}d\Sigma_k=M\,,\qquad
Q[\xi]_H=\frac{\kappa}{2\pi}\frac{(\ell^2 k+r_H^2)}{4G}\int_{\partial \Sigma_H} d\Sigma_k\,.
\end{equation}
Because values of unit areas $\Sigma_k$ for pseudo-sphere, torus, and sphere are $4\pi, 4\pi^2, 4\pi$, respectively, the entropy of these black holes (see \cite{Olea:2005gb}) yields the form
\begin{equation}
   Entropy =\frac{Area}{4G}+ \frac{4\pi \ell^2 k}{4G}\,,
\end{equation} 
so the torus geometry exhibits lack of the shift, and pseudo-sphere can lead to negative entropies.
\subsection{AdS--Kerr}
The geometry of rotating AdS-Kerr black holes can be expressed by the tetrads 
\begin{eqnarray}
e^{0}&=&\frac{\sqrt{\Delta _{r}}}{ \rho }\left(dt-\frac{a}{\Xi}\sin ^{2}\theta d\varphi \right),\qquad e^{1}=\rho
\frac{dr}{\sqrt{\Delta _{r}}},\nonumber\\  
e^{2}&=&\rho \frac{d\theta }{\sqrt{\Delta _{\theta }}},\qquad e^{3}=\frac{\sqrt{\Delta
_{\theta }}}{ \rho }\sin \theta \left(\frac{(r^{2}+a^{2})}{\Xi}d\varphi -a\,dt\right), 
\end{eqnarray}
where we additionally define
$$\rho^{2}=r^{2}+a^{2}\cos ^{2}\theta\, ,\quad \Delta _{r}=(r^{2}+a^{2})\left( 1+\frac{r^{2}}{l^{2}}\right) -2MGr\,,$$
$$ \Delta _{\theta }=1-\frac{a^{2}}{l^{2}}\cos ^{2}\theta\,,\quad\Xi =1-\frac{a^{2}}{l^{2}}\,.$$
For the Killing vector being horizon generator ($\xi = \xi_{t} + \Omega_H \xi_{\varphi}$) we need the angular velocity 
\begin{equation}\label{e43}
    \tilde \Omega =-\frac{g_{t\varphi}}{g_{\varphi\varphi}}=\frac{a\Xi\left( \Delta_\theta(r^2+a^2)-\Delta_r\right)}{ (r^2+a^2)^2\Delta_\theta-a^2\Delta_r \sin^2\theta}
\end{equation}  
evaluated at the horizon (defined by the largest zero of $\Delta_{r}$):
\begin{equation}
    \Omega_H =\frac{a\left(1-\frac{a^2}{\ell^2} \right)}{r^2_H +a^2}\,.
\end{equation}
Besides $\Omega_H$ we will later need the value of (\ref{e43}) at the infinity: $\Omega_\infty=-a/l^{2}$. 

Although this time components of (\ref{e28}) corresponding to the Holst and Pontryagin modification are not just zero anymore, the whole impact related to the Immirzi parameter, once again, drops out. The extensive summation over elements of $\omega^{ab}_{t}$, $\omega^{ab}_{\varphi}$, as well as $I_{\xi} \omega^{ab} = (\omega^{ab}_{t} + \Omega_H\omega^{ab}_{\varphi})$ gives rise to complicated expressions, but at the end, the whole Immirzi contribution is canceled out by the integration at the specified boundaries. Thus, the Noether charges for the Killing vectors associated with the time and rotational invariance stay in an agreement with \cite{Olea:2005gb} and read, respectively, as 
\begin{equation}
Q\left[\frac{\partial }{\partial t}\right]=\frac{M}{\Xi}\,,\qquad
Q\left[\frac{\partial
}{\partial \varphi }\right]=\frac{Ma}{\Xi ^{2}}\,.
\end{equation}
Second expression is the angular momentum $J$, but the first quantity cannot be regarded as the mass for the Kerr-AdS black hole. As it was pointed out by Gibbons, Perry and Pope \cite{Gibbons:2004ai}, it is because the Killing field $\partial _{t}$ is still rotating at radial infinity. The non-rotating timelike Killing vector is expressed by the combination $\partial _{t}-\left( a/l^{2}\right) \partial _{\varphi }$, that substituted in the charge formula gives the physical mass 
\begin{equation}
Mass=Q\left[\partial _{t}-\frac{a}{l^{2}}\partial _{\varphi }\right] =\frac{M}{
\Xi ^{2}}\,.
\end{equation}
Also the angular velocity in the first law of black hole thermodynamics $dM=TdS+\Omega \,dJ$ should be measured relative to a frame non-rotating at infinity:
\begin{equation}
    \Omega =\Omega_H-\Omega_\infty=\frac{a\left(1+\frac{r^2_H}{\ell^2} \right)}{r^2_H +a^2}\,.
\end{equation}
To complete this analysis we finally write formulas for the surface gravity and the entropy
\begin{equation}
    \kappa=\frac{r_H \left(\frac{a^2}{l^2}-\frac{a^2}{r_H^2}+\frac{3 r_H^2}{l^2}+1\right)}{2 \left(a^2+r_H^2\right)}\,,\qquad     Entropy=\frac{4\pi  \left(r^2_H+a^2\right)}{4G \left(1-\frac{a^2}{l^2}\right)}+\frac{4\pi  l^2}{4G}\,,
\end{equation}
to find out that the entropy is again exactly of the form (\ref{entropy1}).

\section{Immirzi parameter and the off-diagonal condition}

In spite of the presence of the Immirzi parameter in the generalized formula (\ref{e28}), the resulting thermodynamics analyzed so far does not contain any trace of it. Let us try to find the reason for this disappearance by coming back to the Noether charge for our action, but now writing it strictly in the metric formulation. For the axisymmetric stationary spacetime with the Killing vector $\partial_\chi$, being either $\partial_t$ or $\partial_\varphi$, and remembering the definition of the connection $\omega^{ab}_\chi=e^{\nu\,a}\nabla_\chi
e^b_\nu=e^{\nu\,a}\left(\partial_\chi
e^b_\nu-\Gamma^\lambda_{\;\chi\nu}e^{b}_\lambda \right)$, it is given by
\begin{align}
 Q[\partial_\chi]&=\frac{2}{32\pi G}\int_{\partial\Sigma} \left(\epsilon^{\mu\nu}_{\quad\theta\varphi}\Gamma_{\mu \chi\nu}-\gamma\,(\Gamma_{\theta \chi\varphi}-\Gamma_{\varphi \chi\theta})\right)\nonumber\\
&+\frac{\ell^2}{32\pi G}\int_{\partial \Sigma} 
   \left(\epsilon^{\mu\nu \rho\sigma }R_{\rho\sigma\theta\varphi}\Gamma_{\mu \chi\nu}-2\gamma  R^{\mu\nu}_{\;\;\;\theta\varphi}\Gamma_{\mu \chi\nu}\right)
\end{align}
Further calculations for $\partial_t$ and the Holst part alone lead to an interesting condition
\begin{equation}\label{Holst-tribute}
-\gamma\,\int_{\partial \Sigma}(\Gamma_{\theta t\varphi}-\Gamma_{\varphi t\theta})= \gamma  \int_{\partial \Sigma}\partial_\theta g_{t\varphi}\,,
\end{equation}
which strongly suggests exploring the non-diagonal metrics. Yet, for most obvious choice being the AdS--Kerr metric we cannot achieve the goal because the non-zero expressions under integrals are canceled by the integration limits at the horizon and at infinity. 

The author of \cite{Yu:2007zze} also notices that the Holst term contributes to the entropy formula derived from the Noether charge expression (through so-called dual horizon area), but claims that its contribution always drops out for the stationary systems, and one should turn to the dynamical black holes to observe demanded influence. 

As we will see in the next section, in spite of this claim, we can find the condition (\ref{Holst-tribute}) fulfilled by the class of an exact Einstein's solution called the AdS--Taub--NUT spacetimes, which exhibits highly nontrivial modification of thermodynamics.

The same conclusion about $\gamma$ and off-diagonal metrics, in particular the Taub--NUT spacetimes, can be drawn by looking at the Holst surface term directly in the path integral formulation \cite{Liko:2011cq}. 

\section{AdS–-Taub-–NUT and Immirzi modifcation of the mass and entropy}
The Taub--NUT spacetime \cite{Misner:1963fr}, introduced by Taub, Newman, Unti and Tamburino, is the generalization of Schwarzschild metric carrying the NUT charge $n$ being gravitational analog of the magnetic monopole. This generalization to the AdS black holes with a NUT charge in 3+1 dimension
$$
    e^0=f(r)\, dt+2  n f(r) \cos\theta\, d\varphi\,\quad e^1=\frac{1}{f(r)} dr\,,$$
$$
e^2= \sqrt{n^2+r^2} \, d\theta\,,\quad e^3= \sqrt{n^2+r^2}\, \sin\theta\,d\varphi, 
$$
translates to the metric   
\begin{equation}
 ds^2=-f(r)^2(dt+2  n \cos\theta\, d\varphi)^2+\frac{dr^2}{f(r)^2}+(n^2+r^2)\,d\Omega^2
\end{equation}
with
\begin{equation}
    f(r)^2= \frac{r^2-2 GMr-n^2+(r^4+6 n^2 r^2-3n^4)\,\ell^{-2}}{n^2+r^2}\,,
\end{equation} 
clearly restoring the AdS--Schwarzschild solution in $n\to 0$ limit. 

This metric is problematic in many ways; it was even called by Misner a "counterexample to almost anything" in General Relativity. It has no curvature singularities, but for $\theta=0$ and $\theta=\pi$ metric fails to be invertible. It gives rise to the Misner string \cite{Misner:1963fr, Mann:1999pc, Clarkson:2005mc} being the gravitational analogue of the Dirac string. To deal with this, one has to impose a periodicity condition ensuring the metric regularity. This forces the relation between the horizon radius $r_H$ and charge $n$, which leads to two separate Euclidean systems called the Taub-NUT and the Taub-Bolt solution \cite{Mann:2004mi, Fatibene:1999ys, Lee:2008yqa}. 

We will follow now only straightforward and naive evaluation of the formula (\ref{e28}) in the Lorentzian regime, leaving rigorous setting and analysis for the future paper. To this end, the Noether charge calculated for the Killing field $\partial_t$ yields the form 
\begin{align}
 Q[\partial_t]&= \frac{  f(r) f'(r)\left(n^2+r^2\right)}{2 G}+ \gamma\;\frac{  n  f(r)^2}{2 G}\nonumber\\
&+\frac{\ell^2}{2G}\left( f(r) f'(r)\left(1+\frac{5n^2-r^2}{n^2+r^2}\,f(r)^2\right)-\frac{2rn^2f(r)^4}{(r^2+n^2)^2}\right)\nonumber\\
&+ \gamma \,\frac{\ell^2\,n}{2G}\left(-2\left(f(r)f'(r)\right){}^2+\frac{2r f(r)^2}{r^2+n^2}f(r)f'(r)+\frac{f(r)^2}{r^2+n^2}\left( 1+\frac{3n^2-r^2}{r^2+n^2}f(r)^2\right) \right)
\end{align}
containing the contributions from Einstein-Cartan, Holst, Euler, and Pontryagin terms. 

We find that the mass obtained from it, is appended by the term affected by the Immirzi parameter
\begin{equation}
    Mass=M+\gamma\;\frac{n \left(\ell^2+4 n^2\right)}{G \ell^2}\,.
\end{equation}
At infinity the Holst term is responsible for $n(1/2+
3n^2/\ell^2)/G$, where the Pontryagin for $n(1/2+ n^2/\ell^2)/G$, with the precise cancellations of divergent terms between them. Surprisingly, there is an intriguing coincidence between $\gamma$ addition above and the mass obtained from the Taub-NUT solution \cite{Chamblin:1998pz} coming from the periodicity conditions for the Euclidean theory.

Let us now turn to the horizon defined as usual by $f(r_H)^2=0$ with the surface gravity being equal $\kappa=f'(r_H)f(r_H)$. Noether charge for the part without topological terms
\begin{equation}
    Q[\partial_t]_{Einstein+Holst}=\frac{  f(r) f'(r)\left(n^2+r^2\right)}{2 G}+\frac{ \gamma  n  f(r)^2}{2 G}
\end{equation}
allows us to observe that Holst contribution to the entropy drops by the horizon definition. Nevertheless, the parameter $\gamma$ eventually appears in the entropy due to the Pontryagin term, which adds value proportional to the surface gravity expression \cite{Kouwn:2009qb}
\begin{equation}\label{surface}
    \kappa=\left(\frac{1}{2} \frac{\partial f(r)^2}{\partial r}
\right)\Big|_{r_H}=\frac{1}{2}\left(\frac{1}{r_H}+\frac{3(n^2+r_H^2)}{\ell^2\, r_H}  \right)
\end{equation}
placed inside the bracket of the formula 
\begin{equation}
    Q[\partial_t]_H= \frac{\kappa}{2\pi} \frac{4\pi \Big(r_H^2+n^2+\ell^2 (1-2 n \gamma  \kappa )\Big)}{4 G}\,.
\end{equation}
The final outcome contains $Area=4\pi(r_h^2+n^2)$, cosmological part $4\pi\ell^2$,  and the extra term
\begin{equation}
 Entropy=\frac{Area}{4 G}+\frac{4\pi \ell^2}{4 G}- \gamma\,\frac{2\pi n\ell^2  }{G}\,\kappa\,.
\end{equation}
Notice, that we are still in Lorentzian regime and no periodicity had been imposed, so there is no condition relating temperature to $1/(8 \pi n)$. Therefore, with the help of (\ref{surface}), we can establish
\begin{equation}
 Entropy=\frac{Area}{4 G}\left(1-\gamma\,\frac{3 n}{r_H}\right)+\frac{4\pi \ell^2}{4 G}\left(1-\gamma\,\frac{ n}{r_H}\right),
\end{equation}
which is the main result of this section. Still it is not clear if this spacetime is too  pathological, which results in this inconsistency, or this should be treated as a premise of  change of the first law of the black hole thermodynamics to take care off the change of a NUT charge. 

\section{Discussion: what about the Immirzi parameter?}
Wald's approach for first order gravity requires regularization procedure to obtain finite charges. Remedy in the form of the MacDowell-Mansouri formulation of gravity allows for straightforward $BF$ theory generalization to the description of black hole thermodynamics with the Immirzi parameter. The generalized formula derived from the $SO(2,3)$ $BF$ theory action, for which topological terms are essential to secure finite charges and having well defined action principle, seems to offer formally different kind of modification, than the one known from the LQG framework. Moreover, in the explicit results for the most common AdS cases (Schwarzschild, Kerr, topological black holes) we find no trace of the desired contribution, each time facing cancellations from the both Holst and Pontryagin terms. 

The aberration appears only for the NUT charged spacetimes, where the Immirzi parameter has an impact not only on the entropy but also on the total mass, which is quite important and interesting result. In the analysis presented above parameter $\gamma$ is always coupled to the NUT charge $n$, so when $n$ is going to zero we lose the whole modification. We also report that entropy is not changed by the Holst term but by the Pontryagin term we have added to the action to avoid divergent charges at infinity, and assure well defined variation principle. Finally we point out that these results seem to agree with those obtained using Euclidean path integrals \cite{Liko:2011cq}.
Description presented above is far from being complete. It is just restricted to straight evaluation and does not properly handle the Misner string. Careful analysis in a much wider context (especially related to the discussions carried in \cite{Hawking:1998ct}, \cite{Clarkson:2002uj}, \cite{Astefanesei:2004ji}, and \cite{Holzegel:2006gn}), will be one of the goals of the forthcoming publication.

\chapter{AdS-Maxwell: (super) algebra and (super) gravity}

\def\bbone{{\mathchoice {\rm 1\mskip-4mu l} {\rm 1\mskip-4mu l}
{\rm 1\mskip-4.5mu l} {\rm 1\mskip-5mu l}}}

\section{Introduction and motivation to Maxwell algebra}

The Maxwell symmetry is an extension of Poincar\'e symmetry arising when one considers symmetries of fields evolving in flat Minkowski space in the presence of a constant electromagnetic background \cite{Bacry:1970ye}, \cite{Schrader:1972zd}. A well known theorem does not allow for central extension of Poincar\'e algebra (see e.g., \cite{Galindo:1967}, \cite{Soroka:2011tc}, \cite{Gomis:2009dm}). The Maxwell algebra is a non-central extension obtained by replacing the commutator of translations $[P_a, P_b]=0$
with
\begin{equation}
    [\mathcal{P}_a, \mathcal{P}_b] = -i \mathcal{Z}_{ab}\, .
\end{equation}
Six new generators $\mathcal{Z}_{ab}=-\mathcal{Z}_{ba}$ commute with themselves and
translations, and follow
\begin{equation}    [\mathcal{M}_{ab},\mathcal{Z}_{cd}]=-i(\eta_{ac}\mathcal{Z}_{bd}+\eta_{bd}\mathcal{Z}_{ac}-\eta_{ad}\mathcal{Z}_{bc}-\eta_{bc}\mathcal{Z}_{ad})\,.
\end{equation}
The Maxwell symmetry did not attract much interest, which seems surprising, because physical systems with constant electromagnetic field, are frequently encountered in physics. 

Recently it was argued in \cite{deAzcarraga:2010sw} that by making use of the gauged Maxwell algebra one can understand it as a source of an additional contribution to the cosmological term in Einstein gravity. However, contrary to the method of constructing the action only locally Lorentz invariant presented by Azcarraga, Kamimura, and Lukierski, we will discuss a different reasoning by demanding an invariance also due to the Maxwell symmetry. This step involves using the $BF$ theory, which easily incorporates such algebraic modification. Because our model requires the AdS algebra to work, first we will have to find the AdS counterpart of the Maxwell algebra. After achieving gravity based on this new algebra, we will perform its supersymmetric extension, which will be rather straightforward after Chapter IV, where we have collected the all relevant definitions and formulas.

\section{AdS-Maxwell algebra}
It turns out that the AdS-Maxwell algebra has the form of a direct
sum of the Lorentz and anti de Sitter algebras, $so(1,3)\oplus so(2,3)$. Thus, the AdS-Maxwell algebra with the generators $\mathcal{P}_{a}$, $\mathcal{M}_{ab}$, and $
\mathcal{Z}_{ab}$ satisfies following commutational relations
\begin{align}\label{c7s-3}
[\mathcal{P}_{a},\mathcal{P}_{b}]&=i(\mathcal{M}_{ab}-\mathcal{Z}_{ab})\,
,\nonumber\\
[\mathcal{M}_{ab},\mathcal{M}_{cd}]&=-i(\eta_{ac}\mathcal{M}_{bd}+\eta_{bd}\mathcal{M}_{ac}-\eta_{ad}\mathcal{M}_{bc}-\eta_{bc}\mathcal{M}_{ad})
,\nonumber\\
[\mathcal{M}_{ab},\mathcal{Z}_{cd}]&=-i(\eta_{ac}\mathcal{Z}_{bd}+\eta_{bd}\mathcal{Z}_{ac}-\eta_{ad}\mathcal{Z}_{bc}-\eta_{bc}\mathcal{Z}_{ad}),\\
[\mathcal{Z}_{ab},\mathcal{Z}_{cd}]&=-i(\eta_{ac}\mathcal{Z}_{bd}+\eta_{bd}\mathcal{Z}_{ac}-\eta_{ad}\mathcal{Z}_{bc}-\eta_{bc}\mathcal{Z}_{ad}),\nonumber\\
[\mathcal{M}_{ab},\mathcal{P}_c]&=-i(\eta_{ac}\mathcal{P}_{b}-\eta_{bc}\mathcal{P}_{a}),\nonumber\\
[\mathcal{Z}_{ab},\mathcal{P}_{c}]&=0\,.\nonumber
\end{align}
Based on them we define the connection gauge field as
\begin{equation}\label{c7-8}
\mathbb{A}_{\mu}=\frac{1}{2}\omega_\mu{}^{ab}
\mathcal{M}_{ab}+\frac{1}{\ell}e_\mu{}^{a}
\mathcal{P}_{a}+\frac{1}{2}h_\mu{}^{ab} \mathcal{Z}_{ab}
\end{equation}
and its curvature $$\mathbb{F}_{\mu\nu}=\partial_\mu\mathbb{A}_{\nu}-\partial_\nu\mathbb{A}_{\mu}-i[\mathbb{A}_{\mu},\mathbb{A}_{\nu}]\,.
$$
It's now decomposed into Lorentz, translational, and Maxwell parts
\begin{equation}\label{c7-10}
\mathbb{F}_{\mu\nu}=\frac12\, F_{\mu\nu}^{ab}\, \mathcal{M}_{ab}
+\frac{1}{\ell}T_{\mu\nu}^{a}\,\mathcal{P}_{a}+ \frac12\,
G_{\mu\nu}^{ab}\, \mathcal{Z}_{ab}
\end{equation}
with a new curvature associated with the generator $\mathcal{Z}_{ab}$
\begin{eqnarray}
G_{\mu\nu}^{ab}&=&D^\omega_\mu h^{ab}_\nu -D^\omega_\nu
h^{ab}_\mu-\frac{1}{\ell^2}( e^a_\mu e^b_\nu- e^a_\nu e^b_\mu)
+(h^{ac}_\mu h^{\quad b}_{\nu\,c}-h^{ac}_\nu h^{\quad
b}_{\mu\,c})\label{c7-13}\,,
\end{eqnarray}
$$G^{ab}=\frac{1}{2}G_{\mu\nu}^{ab}\,dx^\mu\wedge dx^\nu=d h^{ab}+\omega^{ac}\wedge  h^{\;\;b}_{c}+\omega^{bc}\wedge  h_{\;\;c}^{a}-\frac{1}{\ell^2} e^a \wedge e^b+h^{ac}\wedge  h^{\;\;b}_{c}\,.
$$ 
For the Bianchi identity we need covariant derivative $D_\mu^{\mathbb A}$ acting on the full curvature $\mathbb{F}_{\mu\nu}$
\begin{equation}\label{c7-14}
 \epsilon^{\mu\nu\rho\sigma}   D_\mu^{\mathbb A} \mathbb
 F_{\nu\rho}(\mathbb{A})=0\, ,
\end{equation}
which can be decomposed into
\begin{eqnarray}
\epsilon^{\mu\nu\rho\sigma} \,D_\mu^{\omega}R_{\nu\rho}^{ab}=0\label{c7-15}\\
\epsilon^{\mu\nu\rho\sigma} \,(D_\mu^{\omega}T_{\nu\rho}^a-R_{\mu\nu}^{ab} e_{\rho b})=0\label{c7-16}\\
\epsilon^{\mu\nu\rho\sigma}\,(D_\mu^{(\omega+h)}G_{\nu\rho}^{ab}+2h_\mu^{ac}
F_{\nu\rho c}{}^{b}-\frac{2}{\ell^2} e_\mu^a
T_{\mu\nu}^b)=0\label{c7-17}
\end{eqnarray}
It is quite important to notice, that the last identity after some manipulations can be rewritten in a more compact form
\begin{equation}\label{c7-18}
\epsilon^{\mu\nu\rho\sigma}\,
D_\mu^{(\omega+h)}(G_{\nu\rho}^{ab}+F_{\nu\rho}^{ab})=\epsilon^{\mu\nu\rho\sigma}\,D_\mu^{(\omega+h)}R_{\nu\rho}^{ab}(\omega+h)=0\,.
\end{equation}
\section{AdS-Maxwell gravity}
Before turning to the gravity action we need an explicit
form of gauge transformations of the components of connection and
curvature. These gauge transformation read
\begin{equation}\label{c7-19}
\delta_\Theta \mathbb{A}_{\mu}=\partial_\mu \Theta
-i[\mathbb{A}_{\mu}, \Theta]\equiv D^{\mathbb{A}}_\mu \Theta\, ,
\qquad \delta_\Theta
\mathbb{F}_{\mu\nu}=i[\Theta,\mathbb{F}_{\mu\nu}]\, ,
\end{equation}
where the gauge parameter $\Theta$ decomposes into parameters of
local Lorentz, translation and Maxwell symmetries
\begin{equation}\label{c7-20}
    \Theta=\frac{1}{2}\lambda^{ab}\mathcal{M}_{ab}+\xi^a \mathcal{P}_a+\frac{1}{2}\tau^{ab}\mathcal{Z}_{ab}\, .
\end{equation}
By direct calculation we see that the connection components
transform as follows
\begin{eqnarray}
        \delta_\Theta h^{ab}_{\mu}&=&D^\omega_\mu \tau^{ab}-\frac{1}{\ell}(e^a_\mu\,\xi^b-e^b_\mu\, \xi^a)
        +
h^{ac}_\mu(\lambda_{c}^{\;\,b}+\tau_{c}^{\;\,b})+h^{bc}_\mu(\lambda^{a}_{\;\;c}+\tau^{a}_{\;\;c})\nonumber\\
      \delta_\Theta \omega^{ab}_{\mu}&=& D^\omega_\mu \lambda^{ab}+\frac{1}{\ell}(e^a_\mu\, \xi^b-e^b_\mu\, \xi^a)
      \label{c7-21}\\
    \frac{1}{\ell}\delta_\Theta e^{a}_{\mu}&=&D^\omega_\mu \xi^a-\frac{1}{\ell}\lambda^a_{\;b} \,e^b_\mu\,,\nonumber
\end{eqnarray}
while for the components of curvature we find
\begin{eqnarray}
        \delta_\Theta G^{ab}_{\mu\nu}&=&\frac{1}{\ell}[\xi,T_{\mu\nu}]^{ab}-[\tau,F_{\mu\nu}]^{ab}-[(\lambda+\tau),G_{\mu\nu}]^{ab}\nonumber\\
      \delta_\Theta F^{ab}_{\mu\nu}&=&-\frac{1}{\ell}[\xi,T_{\mu\nu}]^{ab}-[\lambda,F_{\mu\nu}]^{ab}  \label{c7-22}\\
    \frac{1}{\ell}\delta_\Theta T^{a}_{\mu\nu}&=&-\frac{1}{\ell}\lambda^a_{\;b}T^b_{\mu\nu}+\xi_b\,F^{ab}_{\mu\nu}\,.\nonumber
\end{eqnarray}
Let us now turn to the construction of the AdS-Maxwell analogue of
deformed $BF$ action known from (\ref{ch3-1}). The generalization of the first term in action should look like $$2B_a\wedge T^a + B_{ab}\wedge F^{ab} +
C_{ab}\wedge G^{ab}\,, \quad \mbox{with}\quad B^a=B^{a4}\,.$$
These combination of terms must be invariant under action of all local symmetries of the
theory, and this requirement fixes the transformation rules for the fields $B$ and $C$ to be
\begin{equation}\label{c7-23}
 \delta_\xi  B^{ab}=({B}^{a} \xi^b-{B}^{b} \xi^a)\,, \quad \delta_\xi C^{ab}=0\,,\quad \delta_\xi
 {B}^{a}=(B^{ab}-C^{ab})\xi_b\,;
\end{equation}
\begin{equation}\label{c7-24}
    \delta_\lambda B^{ab}=-[\lambda,B]^{ab}\,,\quad \delta_\lambda  C^{ab}=-[\lambda,C]^{ab}\,,\quad \delta_\lambda
    {B}^{a}=-\lambda^a{}_{b}{B}^{b}\,;
\end{equation}
\begin{equation}\label{c7-25}
    \delta_\tau B^{ab}=-[\tau,C]^{ab} \,,\quad \delta_\tau C^{ab}=-[\tau,C]^{ab}\,,\quad  \delta_\tau
    {B}^{a}=0\,.
\end{equation}
In the next step we must generalize the second and third term in the action. Looking at (\ref{c7-23})--(\ref{c7-25}) we see that there are
two gauge invariant terms quadratic in the fields
$$ C^{ab}\wedge C_{ab}\qquad \mbox{and}\qquad 2B^a\wedge B_a +  B^{ab}\wedge B_{ab} - 2C^{ab}\wedge B_{ab}\,.$$
In the last step we must find the terms that are generalizations of
the third, gauge breaking term in the action. There
are two combinations of terms satisfying this requirement, namely
$$\epsilon^{abcd}C_{ab}\wedge C_{cd}\qquad \mbox{and}\qquad
\epsilon^{abcd}(B_{ab}\wedge B_{cd}-2C_{ab}\wedge B_{cd})\,.$$
Therefore, the Maxwellian analog of action (\ref{ch3-1}) has the form 
\begin{eqnarray}
 16\pi\, S(A,B)&=& \int 2(B^{a4}\wedge F_{a4}-\frac{\beta}{2}B^{a4}\wedge B_{a4}) \nonumber\\
& &  +B^{ab}\wedge F_{ab}- \frac{\beta}{2}B^{ab}\wedge B_{ab}- \frac{\alpha}{4}\epsilon^{abcd} B_{ab}\wedge B_{cd}\nonumber\\
& &+C^{ab}\wedge G_{ab}- \frac{\rho}{2}C^{ab}\wedge C_{ab} - \frac{\sigma}{4}\epsilon^{abcd} C_{ab}\wedge C_{cd}\nonumber\\
& & + \beta  C^{ab}\wedge B_{ab}+ \frac{\alpha }{2}\epsilon^{abcd}
C_{ab}\wedge B_{cd}\,.\label{c7-26}
\end{eqnarray}
By construction this $BFCG$ action is invariant under local Lorentz and
Maxwell symmetries with the translational symmetry being broken
explicitly by the `epsilon' terms.\\
The algebraic $B$ and $C$ field equations take the form
\begin{eqnarray}
\frac1\ell\, T^{a}&=& \beta B^{a}\label{c7-27}\\
G^{ab}&=&\rho C^{ab}+ \frac{\sigma}{2}\epsilon^{abcd} C_{cd}-\beta B^{ab}-\frac{\alpha}{2}\epsilon^{abcd} B_{cd}\label{c7-28}\\
F^{ab}&=& \beta B^{ab}+ \frac{\alpha}{2}\epsilon^{abcd} B_{cd}-\beta
C^{ab}-\frac{\alpha}{2}\epsilon^{abcd} C_{cd}\label{c7-29}
\end{eqnarray}
Using these equations the action (\ref{c7-26}) can be written in the
simpler form
\begin{equation}\label{c7-30}
 16\pi \, S(A,B)= \frac{1}{2}\int \left(B^{ab}\wedge F_{ab} + C^{ab}\wedge G_{ab}+\frac{2}{\beta}\, B^{a4}\wedge
 F_{a4}\right)\,,
\end{equation}
which after substituting the algebraic equations for $B$ and $C$
fields becomes
\begin{eqnarray}
 16 \pi S(\omega,h,e)&=& \int\left( \frac{1}{4}M^{abcd} F_{ab}\wedge F_{cd}-\frac{1}{\beta\ell^2} \,T^a \wedge T_a\right)\nonumber\\
 & &\quad+\int\frac{1}{4}N^{abcd} (F_{ab}+G_{ab})\wedge
 (F_{cd}+G_{cd})\label{c7-31}
\end{eqnarray}
with $M^{abcd}$ given by (\ref{ch3-6}) and
\begin{equation}\label{c7-32}
    N^{abcd}=\frac{(\sigma-\alpha)}{(\sigma-\alpha)^2+(\rho-\beta)^2}\left(\frac{\rho-\beta}{\sigma-\alpha}\delta^{abcd} -\epsilon^{abcd} \right)\,.
\end{equation}
The first line of (\ref{c7-31}) is just our original action for gravity
with negative cosmological constant (and with Holst and topological
terms). Because our full action has more symmetries than the one considered in \cite{deAzcarraga:2010sw}, the dynamics it describes is much more
restrictive than the one considered in that paper. In fact, one can see that the second line of (\ref{c7-31}) is just... a topological term, which, in particular, does not contribute to the dynamical field equations. 

This follows from the fact that the sum of two curvatures $F^{ab}$ and $G^{ab}$ is the Riemannian curvature of the sum of two connections
$$
F^{ab}(\omega,e)+G^{ab}(h,e) = R^{ab}(\omega+h)\equiv
d(\omega+h)^{ab} + (\omega+h)^{a}{}_c\wedge(\omega+h)^{cb}\, ,
$$
and, in particular the tetrad terms cancel out in this expression.
Therefore the term in the second line of (\ref{c7-31}) is only a sum of
the Euler and Pontryagin invariants, calculated for the connection
$(\omega+h)$. Thus we see that our construction leads just to the Einstein-Cartan
gravity action with the gauge field associated with the Maxwell
symmetry not influencing the dynamics and contributing only to the
boundary terms. In particular the Maxwell terms do not contribute to
the cosmological constant term and we do not see any trace of the
\textit{generalized cosmological term} described in \cite{deAzcarraga:2010sw}. 

Recently the local AdS-Maxwell symmetry was applied to the construction of action being not geometrical \cite{Durka:2011va}. Such a theory has the form of Einstein gravity coupled to the $SO(1,3)$ Yang-Mills fields, and it is both local Lorentz and Maxwell invariant. By this we understand the Einstein-Cartan action (effectively for the $\omega(e)$) appended with the two kinds of Yang-Mills terms for the gauge field $\omega+h$ motivated by the way how Maxwellian symmetry appeared in this chapter. Now, however, due to the Hogde operator contracting spacetime indices, such action will contain dynamical $h$ fields. This form allow us to make a contact with so called $f-g$ and bimetric theories. Since the cosmological model, being the simplest setting, leads in this case to the pathological behavior with a negative energy we will stop its further discussion, and now turn to the supersymmetrization of the AdS-Maxwell algebra.

\section{AdS-Maxwell superalgebra}

Supersymmetrization of the Maxwell algebra leads to a new form of the supersymmetry $\mathcal N=1$, $D=4$ algebra, containing the super-Poincar\'e algebra as its subalgebra (\cite{Soroka:2006aj}, \cite{Bonanos:2009wy}, \cite{Bonanos:2010fw}, and for the latest review see \cite{Kamimura:2011mq}). Let us now follow the supersymmetry extension of the AdS-Maxwell algebra and resulting supergravity. Due to super--Jacobi identities we find that single super-generator $Q$ is not closing the algebra, and addition of a new charge $\Sigma$ is needed. Therefore the AdS-Maxwell superalgebra, being a supersymmetric extension of
(\ref{c7s-3}) contains two supersymmetric generators $Q_\alpha$ and
$\Sigma_\alpha$, both being Majorana spinors with the following
(anti) commutational rules 
\begin{align}\label{c7s-4}
[\mathcal{M}_{ab}, Q_\alpha] &= - \frac{i}2\, (\gamma_{ab}\,
Q)_\alpha\,,\nonumber\\
 [\mathcal{M}_{ab}, \Sigma_\alpha] &=  -\frac{i}2\, (\gamma_{ab}\, \Sigma)_\alpha\,, \nonumber\\
  [\mathcal{Z}_{ab}, Q_\alpha] &=-\frac{i}2\, (\gamma_{ab}\,
  \Sigma)_\alpha\,,\nonumber\\
   [\mathcal{Z}_{ab}, \Sigma_\alpha] &=-\frac{i}2\, (\gamma_{ab}\, \Sigma)_\alpha\,, \nonumber\\
    [\mathcal{P}_a, Q_\alpha]&= -\frac{i}{2}\, \gamma_a\,( Q_\alpha-
    \Sigma_\alpha)\,,\nonumber\\
     [\mathcal{P}_{a}, \Sigma_\alpha] &=0,\nonumber\\
\{Q_\alpha, Q_\beta\}& =-\frac{i}{2}(\gamma^{ab})_{\alpha\beta}\,\mathcal{M}_{ab}+
i(\gamma^a)_{\alpha\beta}\, \mathcal{P}_a\, ,\nonumber\\
\{Q_\alpha,\Sigma_\beta\}
&=-\frac{i}{2}(\gamma^{ab})_{\alpha\beta}\,\mathcal{Z}_{ab}\,,\nonumber\\
\{\Sigma_\alpha, \Sigma_\beta\}
&=-\frac{i}{2}(\gamma^{ab})_{\alpha\beta}\,\mathcal{Z}_{ab}\,,
\end{align}
where all convention from Chapter IV still holds. This superalgebra was achieved in \cite{Durka:2011gm}, and recently confirmed in the work \cite{Kamimura:2011mq}, where it appears as one of the realizations (complementary to \cite{Bonanos:2010fw}) of further generalization the super-Maxwell algebra.

By the Wigner-In\"on\"u contraction of the algebra (\ref{c7s-3}) with
rescaled generators
 $\mathcal{P}_a \rightarrow a\, \mathcal{P}_a$,
$\mathcal{Z}_{ab} \rightarrow a^2\, \mathcal{Z}_{ab}$ and going with
$a$ to infinity we obtain the standard Maxwell algebra. As for the
supersymmetric extension (\ref{c7s-4}), we rescale $Q \rightarrow
a^{1/2}\, Q$ and $\Sigma \rightarrow a^{3/2}\, \Sigma$ to obtain the
Maxwell superalgebra.

Let us now turn to gauging the AdS-Maxwell superalgebra (\ref{c7s-3}),
(\ref{c7s-4}). To this end we write down a gauge field, valued in this
superalgebra
\begin{equation}\label{c7s-5}
    \mathbb{A}_{\mu}=\frac{1}{2}\omega^{ab}_{\mu}\mathcal{M}_{ab}+\frac{1}{\ell}e^a_{\mu}\mathcal{P}_a+\frac{1}{2}h_{\mu}^{ab}\mathcal{Z}_{ab}+\kappa\bar\psi^\alpha_{\mu} Q_\alpha+\tilde{\kappa}\bar\chi^\alpha_{\mu}\Sigma_\alpha
\end{equation}
In this formula $\ell$ is a scale of dimension of length necessary
for dimensional reason, because the tetrad $e^a_{\mu}$ is
dimensionless. Similarly $\kappa$ and $\tilde{\kappa}$ are scales of
dimension $\mbox{length}^{1/2}$ included so as to compensate for
the dimension of the spinor fields. The components of the curvature can be written as
\begin{equation}\label{c7s-7}
    \mathbb{F}_{\mu\nu}=\frac12\, F^{(s)}_{\mu\nu}{}^{ab}\, \mathcal{M}_{ab}+ F^{(s)}_{\mu\nu}{}^{a}\, \mathcal{M}_{a}+ \frac12\, G_{\mu\nu}^{(s)ab}\,\mathcal{Z}_{ab}+  \bar{\mathcal F}_{\mu\nu}^\alpha Q_\alpha+ \bar{\mathcal G}_{\mu\nu}^\alpha \Sigma_\alpha\,,
\end{equation}
where the supercurvatures are given by standard parts and
\begin{align}
  G^{(s)}_{\mu\nu}{}^{ab} &=G_{\mu\nu}^{ab}-\tilde{\kappa}\kappa\, (\bar\psi_\mu\gamma^{ab}\chi_\nu+\bar\chi_\mu\gamma^{ab}\psi_\nu)-\tilde{\kappa}^2\, \bar\chi_\mu\gamma^{ab}\chi_\nu\,,\label{c7s-8}
    \end{align}
with the new pure bosonic curvature
\begin{align}
G_{\mu\nu}^{ab} &=D^\omega_\mu h^{ab}_\nu -D^\omega_\nu
h^{ab}_\mu-\frac{1}{\ell^2}( e^a_\mu e^b_\nu- e^a_\nu e^b_\mu)
+(h^{ac}_\mu h^{\quad b}_{\nu\,c}-h^{ac}_\nu h^{\quad b}_{\mu\,c})\,
    \label{c7s-9}.
\end{align}
We also introduce the fermionic curvature
\begin{align}\label{c7s-11}
{\mathcal G}_{\mu\nu}&=\tilde\kappa \Big((\mathcal D^\omega_\mu\chi_\nu -\mathcal D^\omega_\nu\chi_\mu) +\frac{1}{4}(h^{ab}_\mu\gamma_{ab}\chi_\nu-h^{ab}_\nu\gamma_{ab}\chi_\mu) \nonumber\\
&+\frac{\kappa}{4\tilde\kappa}(h^{ab}_\mu\gamma_{ab}\psi_\nu-h^{ab}_\nu\gamma_{ab}\psi_\mu)-\frac{1}{2\ell}\frac{\kappa}{\tilde\kappa}\left(e_\mu^a\, \gamma_a\psi_\nu -e_\nu^a\, \gamma_a\psi_\mu\right)\Big)\,.
\end{align}
\section{AdS-Maxwell supergravity}
Having these building blocks we proceed to the construction of
the action of the AdS-Maxwell supergravity. To this end we
generalize the construction $BFCG$ theory, and follow symmetrization procedure from Chapter IV, including an additional 2-form fermionic field $\mathcal{C}^\alpha$ associated with the supercharge $\Sigma_\alpha$. 
\begin{align}
64\pi\mathcal{L} &=\left( B_{\mu\nu}^{IJ} F^{(s)}_{\rho\sigma\,IJ} -
\frac\beta2\, B_{\mu\nu}^{IJ} B_{\rho\sigma\,IJ}
-\frac\alpha4\epsilon_{abcd} B_{\mu\nu}^{ab}B_{\rho\sigma}^{cd}
\right)\epsilon^{\mu\nu\rho\sigma}\nonumber\\
&+\left( C_{\mu\nu}^{ab} G^{(s)}_{\rho\sigma\,ab} - \frac\rho2\,
C_{\mu\nu}^{ab}C_{\rho\sigma\,ab}-\frac\sigma4\epsilon_{abcd}
C_{\mu\nu}^{ab}
C_{\rho\sigma}^{cd}\right)\epsilon^{\mu\nu\rho\sigma}\nonumber\\
&+ \left(\beta\,
C_{\mu\nu}^{ab}B_{\rho\sigma\,ab}+\frac\alpha2\epsilon_{abcd}
C_{\mu\nu}^{ab}
B_{\rho\sigma}^{cd}\right)\epsilon^{\mu\nu\rho\sigma}\nonumber\\
&+4\,\left( \bar{\mathcal B}_{\mu\nu}{\mathcal F}_{\rho\sigma}-
\frac\beta2\,\bar{\mathcal B}_{\mu\nu}{\mathcal B}_{\rho\sigma}-
\frac\alpha2\,\, \bar{\mathcal B}_{\mu\nu}\gamma^5{\mathcal
B}_{\rho\sigma} \right)\epsilon^{\mu\nu\rho\sigma}\nonumber\\
&+4\,\left( \bar{\mathcal C}_{\mu\nu}{\mathcal G}_{\rho\sigma}-
\frac\rho2\,\bar{\mathcal C}_{\mu\nu}{\mathcal C}_{\rho\sigma}-
\frac\sigma2\,\, \bar{\mathcal C}_{\mu\nu}\gamma^5{\mathcal
C}_{\rho\sigma} \right)\epsilon^{\mu\nu\rho\sigma}\,
\nonumber\\
&+4\,\left(\frac{\beta}{2}\,\bar{\mathcal C}_{\mu\nu}{\mathcal
B}_{\rho\sigma}+\frac{\beta}{2}\,\bar{\mathcal B}_{\mu\nu}{\mathcal
C}_{\rho\sigma}+  \frac{\alpha}{2}\, \bar{\mathcal
C}_{\mu\nu}\gamma^5{\mathcal B}_{\rho\sigma} + \frac{\alpha}{2}\,
\bar{\mathcal B}_{\mu\nu}\gamma^5{\mathcal C}_{\rho\sigma}
\right)\epsilon^{\mu\nu\rho\sigma}\,.\label{c7s-12}
\end{align}
The bosonic part of this action coincides with the action of
AdS-Maxwell gravity derived in \cite{Durka:2011nf}, while the action
(\ref{c7s-12}) with $\mathcal{C}=\mathcal{G}=0$ is just the $\mathcal{N}=1$
supergravity action in the constrained $BF$ formalism constructed in earlier and in
\cite{Durka:2009pf}.

Algebraic field equations for the fermionic two form fields give
\begin{equation}
    \mathcal{B}-\mathcal{C}=\frac{1}{\alpha^2+\beta^2}\left(\beta\bbone
-\alpha\,\gamma^5\, \right)\mathcal{F}\,, \quad\mbox{and}\quad
 \mathcal{C}=\frac{(\rho-\beta)\bbone-(\sigma-\alpha)\,\gamma^5}{ (\sigma-\alpha)^{2}+(\rho-\beta)^{2}} \Big( \mathcal{G}+\mathcal{F}
 \Big)\, ,\label{c7s-13}
\end{equation}
which after substituting back to the fermionic part of the action
(\ref{c7s-12}) gives
\begin{align}
16\pi \mathcal
L^{f}&=\epsilon^{\mu\nu\rho\sigma}\frac{\alpha}{(\alpha^2+\beta^2)}\,\bar{\mathcal{F}}_{\mu\nu}\left(
\frac{{\beta}\bbone
-{\alpha}\gamma^5}{2{\alpha}}\right)\,\mathcal{F}_{\rho\sigma} \nonumber\\
&+\epsilon^{\mu\nu\rho\sigma}\frac{(\sigma-\alpha)}{(\sigma-\alpha)^2+(\rho-\beta)^2}\,(\bar{\mathcal{G}}_{\mu\nu}+\bar{\mathcal{F}}_{\mu\nu})\left(
\frac{(\rho-\beta)\bbone
-(\sigma-\alpha)\gamma^5}{2(\sigma-\alpha)}\right)
\,(\mathcal{G}_{\rho\sigma}+\mathcal{F}_{\rho\sigma}) \label{c7s-14}
\end{align}
Similarly for the bosonic part of the action we get (see
\cite{Durka:2011nf} for details)
\begin{align}\label{c7s-15}
16\pi \mathcal
L^{b}&=\epsilon^{\mu\nu\rho\sigma}\left(\frac{1}{\beta}
F^{(s)a4}{}_{\mu\nu}  F_{a4}^{(s)}{}_{\rho\sigma}+\frac{1}{4}
M^{abcd} F_{ab}^{(s)}{}_{\mu\nu}
F_{cd}^{(s)}{}_{\rho\sigma}\right)\nonumber\\ &+
\epsilon^{\mu\nu\rho\sigma}\,\frac{1}{4}N^{abcd}
\left(G_{ab}^{(s)}{}_{\mu\nu}+F_{ab}^{(s)}{}_{\mu\nu}\right)
\left(G_{cd}^{(s)}{}_{\rho\sigma}+F_{cd}^{(s)}{}_{\rho\sigma}\right)
\end{align}
with
\begin{align}
M^{abcd}&=\frac{\alpha}{(\alpha^2+\beta^2)}( \gamma\,
\delta^{abcd}-\epsilon^{abcd}) \,,\nonumber\\
N^{abcd}&=\frac{(\sigma-\alpha)}{(\sigma-\alpha)^2+(\rho-\beta)^2}\left(\frac{\rho-\beta}{\sigma-\alpha}\delta^{abcd}
-\epsilon^{abcd} \right)\label{c7s-15a}
\end{align}
Parameters of the model $\alpha$,
$\beta$, and $\ell$ are related to the physical coupling constants:
Newton's constant $G$, cosmological constant $\Lambda$, and Immirzi
parameter $\gamma$ as was shown in (\ref{ch3-1b}). 

Using (\ref{c7s-14}) and (\ref{c7s-15}) one can check, using the {\em1.5}
formalism, that the action is indeed invariant under the action of
both these local supersymmetries\footnote{More precisely, the
variation of the action is proportional to super-torsion, which
vanishes in the {\em1.5} formalism.}. The action is of course also
invariant under the bosonic symmetries: the local Lorentz and
Maxwell leave it invariant.

After convinced ourselves that the action is invariant we can try
to simplify it. Indeed we see a lot of cancellations taking place.
Since the Maxwell gauge field $h_\mu{}^{ab}$ appears in the bosonic action in the topological
terms and, as a consequence of this, its superpartner $\chi$ should
disappear from the action as well. To see this let us first notice that the curvatures $F^{(s)}$ and
$\mathcal F$ have exactly the same form as in the $\mathcal{N}=1$ AdS
supergravity discussed in Chapter IV and \cite{Durka:2009pf}, so that we must only
consider the  $F^{(s)}+G^{(s)}$ and $\mathcal F+\mathcal G$ terms in
the Lagrangians (\ref{c7s-14}), (\ref{c7s-15}). These terms have the form
\begin{align}\label{c7s-22}
G_{\mu\nu}^{(s)ab}+F_{\mu\nu}^{(s)ab}&=R^{ab}_{\mu\nu}(\omega+h)-\Big(\kappa\bar\psi_\mu+\tilde
\kappa\bar\chi_\mu\Big)\gamma^{ab}\Big(\kappa\psi_\nu+\tilde
\kappa\chi_\nu\Big)\nonumber\\
    \mathcal{G}_{\mu\nu}+\mathcal{F}_{\mu\nu}&=\mathcal D^{(\omega+h)}_\mu \Big(\kappa\psi_\nu+\tilde \kappa\chi_\nu\Big)-\mathcal D^{(\omega+h)}_\nu \Big(\kappa\psi_\mu+\tilde
    \kappa\chi_\mu\Big)\,.
\end{align}
Using these, after some straightforward but tedious calculations, we can bring the Lagrangian to the following form
\begin{align}\label{c7s-23}
  16\pi \mathcal L&=-\left(\frac{\kappa^2}{G}\,\bar\psi_\mu\,\gamma^5\gamma_{ab}\,e^a_\nu\, e^b_\rho+ \frac{2\kappa^2\ell}{G}\,\bar\psi_\mu\,\gamma^5\gamma_{a}\,e^a_\nu\,\mathcal D^\omega_\rho\psi_\sigma\right)\,\epsilon^{\mu\nu\rho\sigma}\nonumber\\
&-\bar\psi_\mu\,\left(\frac{1}{4\beta} \frac{2\kappa^2}{\ell}\gamma_{a}\,T_{\nu\rho}^a+\frac{2\kappa^2\ell}{4G}\,(\gamma\bbone -\gamma^5)\, \gamma_a\,T_{\nu\rho}^a\right)\,\psi_\sigma\,\epsilon^{\mu\nu\rho\sigma}\nonumber\\
&-\frac{1}{4\beta} \left(\frac{1}{\ell^2}T_{\mu\nu}^{a}\,T_{\rho\sigma\,a}+\kappa^4 \,\bar\psi_\mu\gamma^{a}\psi_\nu \, \bar\psi_\rho\gamma_{a}\psi_\sigma\right)\,\epsilon^{\mu\nu\rho\sigma}\\
&+\frac{1}{16} M_{abcd} \left(F_{\mu\nu}^{ab}\,F_{\rho\sigma}^{cd}+\kappa^4\, \bar\psi_\mu\gamma^{ab}\psi_\nu\, \bar\psi_\rho\gamma^{cd}\psi_\sigma\right)\,\epsilon^{\mu\nu\rho\sigma}\nonumber\\
&+\frac{1}{16}N_{abcd}\left(\kappa\bar\psi_\mu+\bar
\kappa\bar\chi_\mu\right)\gamma^{ab}\left(\kappa\psi_\nu+\bar
\kappa\chi_\nu\right) \left(\kappa\bar\psi_\rho+\bar
\kappa\bar\chi_\rho\right)\gamma^{cd}\left(\kappa\psi_\sigma+\bar
\kappa\chi_\sigma\right)\,\epsilon^{\mu\nu\rho\sigma}\nonumber\\
&+\mbox{\em total derivative}\nonumber
\end{align}
Making use of Fierz identities one can check that the last line (not counting total derivatives) in (\ref{c7s-23})
vanishes identically along with other four-fermion terms and there
are some simplifications in the second line. Notice that in this way
there is no trace  of $\chi$ in the bulk Lagrangian anymore. Indeed,
after some cancellations all the $\chi$-dependent terms can be
combined into a total derivative. 

If we set $\kappa^2=4\pi G/\ell$ the Lagrangian reduces exactly to the form (\ref{c3-35}) discussed in Chapter IV, and \cite{Durka:2011nf} with the only change in the total derivative term.

This concludes our construction, in which we showed that the gauged
theory of the AdS-Maxwell supersymmetry is somehow trivial, reducing
just to the standard $\mathcal{N}=1$ supergravity. This is not surprising, since in the bosonic case the gauge field of Maxwell symmetry $h_\mu^{ab}$ appears similarly only through the topological terms. Although these topological terms do not change the bulk field
equations they may influence the asymptotic charges in an
interesting way.\newline

The aim of this chapter was to extend the Maxwell
algebra to the AdS-Maxwell one (for original papers see \cite{Durka:2011nf, Durka:2011gm}),
presenting an alternative construction of the action of
gravity based on the gauging of the AdS-Maxwell algebra employing
the concept of a $BF$ theory \cite{Freidel:2005ak}. 

We find that theory
obtained by this procedure is just the Einstein-Cartan theory with
the additional Holst action term, and a set of topological terms. Field $h^{ab}$ being the gauge field associated with the generators $\mathcal{Z}_{ab}$,
appears only in the topological term that does not influence the
dynamics of the theory. This theory differs therefore from the one
discussed in \cite{deAzcarraga:2010sw}, because there the Maxwell symmetry was not implemented at the level of the construction of the action.

\chapter{Canonical analysis of $SO(4,1)$ constrained BF theory}

\def\SO{\mathsf{SO}}
\def\cP{{\cal P}}
\def\cD{{\cal D}}
\def\cC{{\cal C}}
\def\cL{{\cal L}}
\def\w{\omega}

\section{Canonical analysis}

One of the most important developments in canonical general relativity of the last decades was Ashtekar's discovery that the phase space of gravity can be described with the help of a background independent theory of self-dual connections \cite{Ashtekar:1986yd}. This later became a foundation of the research program of loop quantum gravity \cite{Rovelli:2004tv}, \cite{Thiemann:2007zz}. The original Ashtekar's formulation was generalized few years later by Barbero to the case of real connections \cite{Barbero:1994ap}, parametrized by a single real number $\gamma$, called the Immirzi parameter \cite{Immirzi:1996di}. This leads to the additional term in gravity action called the Holst term.

We are going to show that the considered BF action, now for the de Sitter gauge group (as it was done in a paper \cite{Durka:2010zx}), reflects the set of constraints, and follows the Holst canonical analysis. First step of this analysis for the BF theory defined by (\ref{ch3-1}) is to decompose of the curvature
$F_{\mu\nu}{}^{IJ}$ into two parts $    F_{\mu\nu}{}^{IJ} \rightarrow (F_{0i}{}^{IJ}, F_{ij}{}^{IJ})$, where
\begin{equation}\label{ch7-1}
    F_{0i}{}^{IJ} =\dot A_i{}^{IJ} -\partial_iA_0{}^{IJ}+A_0{}^I{}_K\, A_i{}^{KJ} - A_i{}^I{}_K\, A_0{}^{KJ}=\dot A_i{}^{IJ} - \cD_i A_0{}^{IJ}\,,
\end{equation}
\begin{equation}\label{ch7-2}
    F_{ij}{}^{IJ} =\partial_i A_j{}^{IJ} +A_i{}^I{}_K\, A_j{}^{KJ} - i \leftrightarrow j\, .
\end{equation}
Dot denotes the time derivative, and $\cD_i$ is the
covariant derivative for $A_i{}^{IJ}$. We decompose also the $B$ field
\begin{equation}\label{ch7-3}
    B_{\mu\nu}{}^{IJ} \rightarrow \left(B_{0i}{}^{IJ} \equiv B_{i}{}^{IJ},\, \cP^{i}{}^{IJ}\equiv 2\epsilon^{ijk}\, B_{jk}{}^{IJ}\right)\, .
\end{equation}
Using these definitions and integrating by parts we can rewrite the action as follows
\begin{equation}\label{ch7-4}
   S=\int dt \mathcal{L}\, ,\quad\quad     \mathcal{L}=\int d^3x \big(\cP^{i}{}_{IJ} \dot A_i{}^{IJ} + B_{i}{}^{IJ} \Pi^i{}_{IJ} + A_0{}^{IJ}
    \Pi_{IJ}\big)\, .
\end{equation}
The $\cP^{i}_{IJ}$ turns out to be momenta associated with spacial components of the gauge field $A^{IJ}_i$, while the remaining components of $B_{0i}{}^{IJ}$ play the role of Lagrange multipliers. Also the zero component of the connection  becomes a Lagrange multiplier.
This is enforcing the constraints $\Pi^i{}_{IJ}$ and $\Pi_{IJ}$ to be
\begin{equation}\label{ch7-5}
   \Pi_{IJ}(x) = \left(\cD_i \cP^i\right)_{IJ}(x)
   =\Big(\partial_i\cP^i{}_{IJ} + A_i{}_I{}^K\cP^i{}_{KJ}+A_i{}_J{}^K\cP^i{}_{IK}\Big) (x)\approx 0\,,
\end{equation}
\begin{equation}\label{ch7-6}
   \Pi^i{}_{IJ}(x) = \left(2\epsilon^{ijk}\, F_{jk}{}_{IJ}-\beta\, \cP^{i}_{IJ} - \frac\alpha2\, \epsilon_{IJKL4}\, \cP^i{}^{KL}\right)(x)\approx 0\,.
\end{equation}
The Poisson bracket of the theory is
\begin{equation}\label{ch7-7}
   \left\{ A_i{}^{IJ}(x), \cP^j{}_{KL}(y)\right\}=\frac12\, \delta(x-y)\, \delta_i^j\,
   \delta_{KL}^{IJ}\, .
\end{equation}
The factor $1/2$ results from the fact that the canonical momentum
associated with $A$ is defined as $\delta \mathcal{L}/ \delta \dot A$ is
$2\cP$, not $\cP$.) The Lagrangian (\ref{ch7-4}) contains just the
standard ($p\dot q$) kinetic term appended with a combination of
constraints, reflecting the manifestation of diffeomorphism
invariance of the action (\ref{ch3-1}) that we have started with. It is
worth noticing that prior to taking care of the constraints the
dimension of phase space of the system is $2 \times 3 \times 10 =60$
at each space point. After employing the time gauge, the dimension of the physical
phase space is going to be $4$, as it should be. 

Next steps require tedious and complicated Dirac procedure of a classification of the constraints, dealing with the second class constraints by changing the Poisson bracket to a form of the so called Dirac bracket. For details we send the reader to \cite{Rezende:2009sv}, and \cite{Durka:2010zx}.

For the topological limit $\alpha=0$ all the constraints are first class, but then not all them are independent. Indeed taking the covariant divergence of the $\Pi^i{}_{IJ}$ constraint and making use of the Bianchi identity we see that $(\cD_i \Pi^i)_{IJ}=-\beta \Pi_{IJ}$ and thus the set of constraints is reducible. We have only 30 independent first class constraints $\Pi^i{}_{IJ}$, which remove exactly 60 dimensions from the phase space, as it
should be since the theory with $\alpha=0$ is topological.

Now we will rewrite the constraints (\ref{ch7-5}) and (\ref{ch7-6}) in a form that makes it easier to compare them with the constraints of General Relativity Hamiltonian appended with a Holst term discussed in \cite{Holst:1995pc}. To this end we perform the splitting on the purely Lorentz indices ($\alpha,\beta=0,1,2,3$), and the rest of them with index 4, which below will be explicitly skipped 
\begin{eqnarray}\label{ch7-8}
\Phi^i_\alpha{~}&=&\cP^{i}_{\alpha} - \frac{4}{\ell\beta}\epsilon^{ijk}\, \cD^\omega_j e_k{}_{\,\,\alpha}\approx 0\\
\Phi^i_{\alpha\beta}&=&\cP^{i}_{\alpha\beta}-M_{\alpha\beta}{}^{\gamma\delta}\, F^{}_{jk}{}_{\,\gamma\delta}\,\epsilon^{ijk} \approx 0\\
\Pi_{\alpha\beta}&=& \frac{2}{\ell^2}\,\epsilon^{ijk} \cD^\omega_i \Big(K_{\alpha\beta}{}^{\gamma\delta}\,e_{j\,\gamma}e_{k\,\delta}\Big)\approx 0\\
\Pi_{\alpha}{~}&=& \frac{1}{\ell}\,\epsilon^{ijk}\,K_{\alpha\beta}{}^{\gamma\delta}\, e^{\,\,\beta}_{i}\,R_{jk\,\,\gamma\delta}-\frac{2\alpha}{(\alpha^2+\beta^2)\ell^3}\,\epsilon^{ijk}\,
\epsilon_{\alpha\beta\gamma\delta}\;e^{\;\beta}_{i}\,e^{\;\gamma}_{j}\,e^{\;\delta}_{k}\approx 0
\end{eqnarray}
We have used the operators
\begin{equation}
M^{\alpha\beta}{}_{\gamma\delta}\equiv \frac{\alpha}{(\alpha^2+\beta^2)}( \gamma\, \delta^{\alpha\beta}_{\gamma\delta}-\epsilon^{\alpha\beta}{}_{\gamma\delta}),\qquad
K^{\alpha\beta}{}_{\gamma\delta}\equiv
\frac{\alpha}{(\alpha^2+\beta^2)}(\frac{1}{\gamma}\,
\delta^{\alpha\beta}_{\gamma\delta}+\epsilon^{\alpha\beta}{}_{\gamma\delta})\,
,  
\end{equation}
where find the coupling constants $\alpha$ and $\beta$ satisfying the
identity $\alpha/(\alpha^2+\beta^2)=\ell^2 /G$.

Therefore the Hamiltonian is expressed as a combination of these constraints 
\begin{equation}\label{ch7-9}
    H= -2 A^\alpha\, \Pi_\alpha-A^{\alpha\beta}\, \Pi_{\alpha\beta}-2B_i{}^{\alpha}\, \Phi^i{}_{\alpha}-B_i{}^{\alpha\beta}\, \Phi^i{}_{\alpha\beta}\,.
\end{equation}
To establish the equivalence with the constraint proposed by Holst \cite{Holst:1995pc} we will have to fix the time gauge. But earlier we shall notice that in the case when the constant time surface is without spacial boundaries $\partial \Sigma=0$, the topological terms play the role of the generating functional for canonical transformations,
which simplify the constraints considerably \cite{Rezende:2009sv}.
The key observation is that Pontryagin, Euler and Nieh-Yan invariants can
be expressed as total derivatives. Therefore the topological part of action (\ref{ch3-1}) takes the form
\begin{eqnarray}\label{ch7-10}
    S_T&=& \frac{2\alpha}{(\alpha^2+\beta^2)}\frac{\beta}{\alpha}\int\partial_\mu\Big( \cC ^\mu (^+\omega)+ \cC ^\mu (^-\omega) \Big) -i \frac{2\alpha}{(\alpha^2+\beta^2)}\int\partial_\mu\Big( \cC ^\mu (^+\omega)- \cC ^\mu (^-\omega)\Big)\nonumber \\
&+&\frac{4}{\beta\ell^2}  \int \partial_\mu \big( e_{\nu\;\alpha} \cD^\omega_\rho e^{\;\alpha}_{\sigma} \big)\,\epsilon^{\mu\nu\rho\sigma}\,. 
\end{eqnarray}
In spite of the presence of the imaginary $i$ here, the action $S_T$ is real (for real $\gamma$). For constant time surfaces, being a manifold without boundary ($\partial \Sigma=0$), all total spacial derivatives terms drop out
and only the ones with total time derivative survive $
  S_T=\int \partial_0  W(e,\omega)\, ,
$ where $W(\omega,e)$ is a functional of torsion, and self, and anti-self dual Chern-Simons forms $\cL_{CS}\equiv \cC^0$
\begin{equation}\label{ch7-11}
     W(e,\omega)=\frac{4}{\beta\ell^2} \int_\Sigma  \,\epsilon^{ijk} \, \big( e_{i\;\alpha} \cD^\omega_j e^{\;\alpha}_{k} \big)+\frac{2\alpha}{(\alpha^2+\beta^2)} \int_\Sigma \Big((\gamma-i)\cL_{CS} (^+\omega)+(\gamma+i) \cL_{CS} (^-\omega) \Big)
\end{equation}
Variables of the Hamiltonians, which differ by the time derivative of a functional can be related by the canonical transformation. With the functional $W$ we can make a transformation, which defines the new momenta $\mathscr{P}^i_a$, $\mathscr{P}^i_{ab}$ of the
tetrad $e$ and the connection $\omega$, respectively
\begin{equation}\label{ch7-12}
  \mathscr{P}^i_\alpha=  \cP^i_\alpha+\{\cP^i_\alpha, W(\omega,e)\}, \quad   \mathscr{P}^i_{\alpha\beta}=  \cP^i_{\alpha\beta}+\{\cP^i_{\alpha\beta}, W(\omega,e)\}
\end{equation}
with
\begin{equation}\label{ch7-13a}
 \left\{e_i^\alpha,\mathscr{P}^j_\beta\right\} =\frac{1}{2}\ell\, \delta^j_i\, \delta^\alpha_\beta\quad \mathrm{and}\quad       \left\{\omega_i^{\alpha\beta},\mathscr{P}^j_{\gamma\delta}\right\} =\frac{1}{2}\delta^j_i\,\delta^{\gamma\delta}_{\alpha\beta}\,.
\end{equation}
Since the variations of the functional $W(\omega,e)$ are
\begin{eqnarray}\label{ch7-13}
       \frac{1}{2} \frac{\delta W}{\delta \omega^{\alpha\beta}_{i}}&=&M_{\alpha\beta}{}^{\gamma\delta}\, R^{}_{jk}{}_{\,\gamma\delta}\,\epsilon^{ijk}-\frac{4}{\beta\ell^2}\,e_{j\;\alpha}\, e_{k\;\beta}\,\epsilon^{ijk}~~~~~~\\
    \frac{1}{2}\frac{\delta W}{\delta e_i^\alpha}&=& \frac{4}{\ell\beta}\epsilon^{ijk}\, \cD^\omega_j e_k{}_{\,\,\alpha}
\end{eqnarray}
we find that new constraints, expressed in terms of new momenta (\ref{ch7-12}) are
\begin{eqnarray}
\Phi^i_\alpha{~}&=&  \mathscr{P}^i_\alpha\approx 0,\label{c8-r1x}\\
\Phi^i_{\alpha\beta}&=&\mathscr{P}^i_{\alpha\beta}- \frac{2}{\ell^2}K^{~~~\gamma\delta}_{\alpha\beta} \,e_{j\;\gamma}\,e_{k\;\delta}\,\epsilon^{ijk}\approx 0\label{c8-r1y}\\
\Pi_{\alpha\beta}&=& \frac{2}{\ell^2}\,\epsilon^{ijk}K_{\alpha\beta}{}^{\gamma\delta}\, \cD^\omega_i \Big(e_{j\,\gamma}e_{k\,\delta}\Big)\approx 0\label{c8-r1c}\\
\Pi_{\alpha}{~}&=& \frac{1}{\ell}\,\epsilon^{ijk}\,K_{\alpha\beta}{}^{\gamma\delta}\, e^{\,\,\beta}_{i}\,F_{jk\,\,\gamma\delta}\approx 0\label{c8-r1d}
\end{eqnarray}
In order to make a contact with the Hamiltonian analysis of Holst, we have
to fix the gauge so as to remove the time component of the tetrad
and then to relate momenta associated with Lorentz connection
with an appropriate combination of the remaining tetrad components \cite{Rezende:2009sv}. Therefore we introduce the gauge condition, which must be added to the list of constraints
\begin{equation}\label{ch7-gauge}
    e^0_i \approx 0\,.
\end{equation}
This  leads to a reduction of the constraints removed by the Dirac brackets, and after some redefinitions we can identify the final variable as
 \begin{equation}\label{ch7-14}
  {}^- w^{a}_i =\omega^{0a}_i  -\,  \frac{1}{2\gamma}\,\epsilon^{abc} \omega_{i\;bc}
\end{equation}
for $a=1,2,3$, and similarly identify the momentum of ${}^- w_{j\,b}$ with
$$
-\frac{4\alpha}{(\alpha^2+\beta^2)\ell^2}\,\epsilon^{ijk}\,\epsilon_{abc}\,e^b_j\,e^c_k\, ,
$$
to give the Poisson bracket $\{ {}^- w^{a}_i ,  {}^-\mathds{P}_{b}^j \}=\delta^a_b\delta^j_i$.
What remains is the set of three Gauss $G_a$, three vector $V_a$, and one scalar $S$ constraints, all of them first class, constraining the $18$-dimensional phase space of ${}^- w_{j\,b}$ and its momenta. Thus the dimension of physical phase space is $18-14=4$ as it should. Of course, the final set of constraints we have obtained has exactly the form of the constraints describing gravity in \cite{Holst:1995pc}.

This analysis (for full details see \cite{Durka:2010zx}), although quite involved, seems to be significantly simpler than the analogous one of Plebanski theory reported in   \cite{Buffenoir:2004vx}. It might be relevant to consider spin foam model associated with this particular formulation of gravity. Unfortunately, not much work has been done till now on the $SO(4,1)$ spin foam models, which would require to handle not only the quadratic B field term, but also the representation theory of SO(4,1) group.

%
%
%
%

\chapter{Summary}

Among the many existing methods and strategies for going beyond standard gravity we have concentrated on a deformed topological $BF$ theory, in which the gravity theory emerges as a result of a gauge symmetry breaking. The strength of this model lies in its generality, reaching far beyond of the Einstein's theory. Its number of interesting properties created an excellent opportunity to examine the existing results, and provide a starting point for development and a deeper understanding. The reincarnation of what is known since the late 70's as MacDowell-Mansouri gravity in the form of deformed topological $BF$ theory conceals a rich and interesting structure. Remarkably, its action incorporates all six possible terms, fulfilling all the necessary symmetries (both diffemorphism and local Lorentz invariance) of first order gravity in four dimensions, and is governed only by Newton's gravitational constant, the cosmological constant, and the Immirzi parameter. The resulting structure is, therefore, composed from the Einstein-Cartan action with a negative cosmological constant and a Holst term, appended with three topological terms: Euler, Pontryagin, and Nieh-Yan. Additionally, it has a very intriguing appearance of perturbation theory, in which general relativity is reproduced as a first order perturbation around the topological vacuum related to the unconstrained part of $SO(2,3)$ $BF$ model.  

After presenting the formal structure of this particular model we have followed the most important novelty introduced by it, the consistent description including topological terms and the Immirzi parameter $\gamma$, in order to show that $\gamma$ does not influence  supergravity resulting from the super-BF theory.

Next aim was to generalize the formulas for the gravitational Noether charges (black hole mass, angular momentum, and entropy) to include the Immirzi parameter $\gamma$ using the framework of $BF$ theory introduced by Freidel and Starodubtsev. The outcome generalizes the results achieved in the 70's by Hawking and Bekenstein, et al. 

A finite value of the gravitational charges and, for the AdS asymptotics, automatically well-defined variational principle (both surprisingly assured by the topological terms) can serve as a motivation supporting the investigated model. Achieved results include a shift of the entropy by a constant associated with a cosmological constant, and no contribution in the macroscopic description from the Immirzi parameter in the thermodynamics of the standard (Schwarzschild and Kerr) anti-de Sitter cases. Analyzed thermodynamics emerging from gravitational Noether charges indicates the absence of this parameter for these spaces; it is realized without the standard procedure of fixing $\gamma$ value to remove the prefactor distorting Bekenstein's entropy, like it is done in loop quantum gravity. This leads naturally to a tension between these descriptions, a problem which must be investigated further. Quite surprisingly, the Immirzi parameter has an impact on both the entropy and the total mass in the case of the Taub-NUT spacetime. The change should apply as well to other off-diagonal metrics fulfilling condition (\ref{Holst-tribute}). Full understanding of these results will require additional work and interpretation in the context of LQG, where $\gamma$ modifies the entropy of the black holes even in the standard cases. Hopefully this work brings much to this debate.

Besides these applications we have also investigated gravity and supergravity corresponding to the modified anti-de Sitter algebra (so-called AdS-Maxwell algebra). This symetry alone is motivated by the symmetry of fields in AdS with the constant electromagnetic background. Work done in a collaboration with supervisor J. Kowalski-Glikman and M. Szczachor on this interesting extension of the Poincare/AdS algebra shows that modification of the algebraic structure of this theory is done by the introduction of new generators and fields, and gives some interesting results when applied to gravity. At the same time $BF$ model proved to be very convenient platform to include the Maxwell symmetry and construct (super) gravity.

Main goal of this thesis (being the culmination of the research done during my doctoral studies at the Wroc\l aw Institute for Theoretical Physics under the supervision of Prof. Jerzy Kowalski-Glikman), was to analyze a deformation of topological $BF$ theory as a theory of gravity and supergravity. Several properties of the investigated $BF$ model made it perfect tool to test a wide class of existing results in more general framework. As it turned out, that was a great starting point for pursuing many important subjects and problems concerning formal side of the gravity models, and an intimate relationship between gravity and thermodynamics, along the process preparing well for the new challenges and further exploring fundamental aspects of contemporary physics.

\newpage

%
\appendix
\renewcommand*{\chaptername}{Appendix}
\chapter*{Appendix}
\addcontentsline{toc}{chapter}{Appendix}
Here we present few conventions and formulas that were used in the main text.
\subsection*{p-forms}
Let $A$ be $p$-form, and $B$ $q$-form then
$$ A \wedge B =(-1)^{pq} B \wedge A $$
$$ d(A\wedge B)= dA\wedge B + (-1)^p A\wedge dB$$
where $d$ is exterior derivative
$$ dA=d(A_{k_{1}k_{2}...k_{p}}dx^{k_{1}}\wedge dx^{k_{2}}\wedge...\wedge dx^{k_{p}})=\frac{\partial A_{k_{1}k_{2}...k_{p}}}{\partial x^k}dx^k \wedge dx^{k_{1}}\wedge dx^{k_{2}}\wedge...\wedge dx^{k_{p}}$$

\subsection*{Levi-Civita symbol}
We define $\epsilon^{01234}=\epsilon^{0123}=1$ which means $\epsilon_{0123}=-1$. One also must remember that contraction for the Minkowski reads as
$$\epsilon^{abmn}\epsilon_{mncd}=-(4-2)!\delta^{ab}_{cd}=-2(\delta^a_c \delta^b_d - \delta^a_d \delta^b_c)$$.
\begin{align}
& (\det e^i_\mu)^2=-\det g_{\mu\nu}=-g, & e=\sqrt{-g}\\
& e= \frac{1}{4!}\epsilon_{abcd}\, e_\mu^a e_\nu^b e_\rho^c e_\sigma^d\, \epsilon^{\mu\nu\rho\sigma},& \epsilon^{\mu\nu\rho\sigma}\epsilon_{\mu\nu\rho\sigma}=4!\,\sqrt{-g}
\end{align}
\subsection*{Gamma matrices}
Definition of $\gamma^5=-\gamma_5$ is given by
$$\gamma^5=\gamma^0\gamma^1\gamma^2\gamma^3\qquad  \gamma_5=\gamma_0\gamma_1\gamma_2\gamma_3$$
$$\gamma^5=-\frac{1}{4!}\epsilon_{abcd}\gamma^a\gamma^b\gamma^c\gamma^d\qquad  \gamma_5=\frac{1}{4!}\epsilon^{abcd}\gamma_a\gamma_b\gamma_c\gamma_d$$
It is easy to show that
$\gamma_0^2=\gamma_5^2=-1$ and $\gamma^2_1=\gamma^2_2=\gamma^2_3=1$).\\

Very useful are also other identities and definitions
\begin{eqnarray}
\gamma^{ab}&=&\frac{1}{2}\epsilon^{abcd}\gamma_{cd}\gamma_5,\qquad\gamma_{ab}=-\frac{1}{2}\epsilon_{abcd}\gamma^{cd}\gamma^5\\
\gamma_c\gamma_{ab}&=&\eta_{ca}\gamma_b-\eta_{cb}\gamma_a-\epsilon_{abcd}\gamma^{d}\gamma^5\\ 
\gamma_{ab}\gamma_c&=&\eta_{cb}\gamma_a-\eta_{ca}\gamma_b-\epsilon_{abcd}\gamma^{d}\gamma^5
\end{eqnarray}
\subsection*{Fierz identities}
Relevant Fierz identity:
\begin{equation}
\bar{\psi}_\mu\,\Gamma^A\,\psi_\nu\epsilon^{\mu\nu\rho\sigma}=0\qquad\mathrm{where}\qquad \Gamma^A=\{1,\gamma^5, \gamma^5\gamma^a\}
\end{equation}
Because $    \bar{\psi}\chi= \bar{\chi}\psi\qquad  \bar{\psi}\,\gamma_5\,\chi= \bar{\chi}\,\gamma_5\,\psi\qquad  \bar{\psi}\,\gamma_5\gamma_i\,\chi= \bar{\chi}\,\gamma_5\gamma_i\,\psi$\\
 
We need also another one for arbitrary combination of gamma's:
\begin{equation}
(\bar{\psi}_\mu\,\Gamma\,\psi_\nu)\,(\bar\psi_\rho\, \Gamma\,\psi_\sigma)\,\epsilon^{\mu\nu\rho\sigma}=0
\end{equation}

\subsection*{Covariant derivative}
The covariant derivative acting on spinors:
\begin{eqnarray}
    \mathcal D_{\mu}\bar{\psi}_\nu&=& \partial_{\mu}\bar{\psi}_\nu-\frac{1}{4}\omega^{ab}_\mu\,\bar{\psi}_\nu\,\gamma_{ab}-\frac{1}{2\ell} e^{a}_{\mu}\,\bar{\psi}_\nu\,\gamma_{a}\nonumber\\
\mathcal D_{\mu}\psi_\nu&=& \partial_{\mu}\psi_\nu+\frac{1}{4}\omega^{ab}_\mu\,\gamma_{ab}\,\psi_\nu+\frac{1}{2\ell} e^{a}_{\mu}\,\gamma_{a}\,\psi_\nu\, .
\end{eqnarray}

} 

%
%

\begin{singlespace}

\end{singlespace}




\backmatter


\end{document}

%% file: Abstract.tex
~~

In this thesis we study aspects of gravity seen as a deformation of topological BF theory \cite{Freidel:2005ak}. We start by introducing the model, which originates in the construction known from the 70's as MacDowell-Mansouri gravity \cite{MacDowell:1977jt} combining together the two independent variables in Cartan theory: the spin connection and the tetrad. This intriguing formulation turns out to be the Yang-Mills theory with the incorporated constraint reduced to torsionless condition. Its new reincarnation in the form of the deformed BF model touches interesting aspects of gravity developed in the context of loop quantum gravity, being in some sense successor of the canonical formulation. Structure behind it turns out to be the maximal generalization of gravity in the first order formulation. Considered framework provides the most general form of action, containing the Einstein-Cartan action with a negative cosmological constant, the Holst term and the topological Euler, Pontryagin and Nieh-Yan invariants.

First part of this thesis includes introduction to the construction and formal advantages of the BF model introduced few years ago by Freidel and Starodubtsev \cite{Freidel:2005ak}.

The second part is based on a conduct of six articles \cite{Durka:2009pf, Durka:2011yv, Durka:2011zf, Durka:2011nf, Durka:2011gm, Durka:2010zx} written during doctoral research process with the analysis of various applications. Unique characteristic of the Freidel--Starodubtsev BF model, as well as its generality, makes it very useful tool suitable to follow important questions corresponding to supergravity, black hole thermodynamics, canonical analysis, and AdS algebra modifications. Notable example is the relevance of the Immirzi parameter $\gamma$, and the corresponding Holst modification, as well as the presence of topological terms (Euler, Pontryagin, Nieh-Yan) in any of these topics. 

In a few chapters we will guide the reader through the research and present the outcome of the thesis, with key results being:
\begin{itemize}
    \item lack of influence of the Immirzi parameter, and the corresponding Holst modification, on $\mathcal{N}=1$ supergravity resulting from the super-BF theory
    \item derivation of a consistent description of black hole thermodynamics based on BF theory with the essential role of topological terms in a regularization, and assuring the finite Noether charges for the asymptotically AdS black hole spacetimes 
    \item generalization of the gravitational Noether charges to the presence of the Immirzi parameter 
    \item no trace of the Immirzi parameter in the standardly explored black hole thermodynamics of the AdS-Schwarzschild and AdS-Kerr spacetimes derived from the generalized gravitational Noether charges
    \item revealing the impact of the Immirzi parameter on the black hole mass, and the entropy in the class of NUT charged spacetimes
    \item finding the modification of the anti-de Sitter algebra to the form of the AdS-Maxwell algebra (and then a corresponding extension to the superalgebra) arising from symmetries of systems evolving in a constant electromagnetic background on the AdS space
    \item construction of gravity and supergravity based on the AdS-Maxwell algebra and superalgebra within the BF theory framework
    \item performing Dirac's canonical analysis for the de Sitter case with the topological terms playing the role of canonical transformation, preserving earlier results obtained in the Holst analysis.
\end{itemize}

%% file: Publications.tex

1.	{\color{maroon} 
Supergravity as a constrained BF theory.}\\
R. Durka, J. Kowalski-Glikman, M. Szczachor\\
Published in Phys.Rev. D81 (2010) 045022; arXiv:0912.1095 [hep-th]
\\[10pt]
2.	{\color{maroon} 
Gravity as a constrained BF theory: Noether charges and Immirzi parameter.}\\
R. Durka, J. Kowalski-Glikman\\
Published in Phys.Rev. D83 (2011) 124011; arXiv:1103.2971 [gr-qc]
\\[10pt]
3.	{\color{maroon} 
Immirzi parameter and Noether charges in first order gravity.}\\
R. Durka\\
Published in J.Phys.Conf.Ser. 343 (2012) 012032; arXiv:1111.0961 [gr-qc]
\\[10pt]
4.	{\color{maroon} 
Gauged AdS-Maxwell algebra and gravity.}\\
R. Durka, J. Kowalski-Glikman, M. Szczachor\\
Published in Mod.Phys.Lett. A26 (2011) 2689-2696; arXiv:1107.4728 [hep-th]
\\[10pt]
5.	{\color{maroon} 
AdS--Maxwell superalgebra and supergravity.}\\
R. Durka, J. Kowalski-Glikman, M. Szczachor\\
Published in Mod.Phys.Lett. A27 1250023 (2012); arXiv:1107.5731 [hep-th]
\\[10pt]
6.	{\color{maroon} 
Hamiltonian analysis of SO(4,1) constrained BF theory.}\\
R. Durka, J. Kowalski-Glikman\\
Published in Class.Quant.Grav. 27 (2010) 185008; arXiv:1003.2412 [gr-qc]
\newpage 
\begin{center}
    \textbf{Abstracts}
\end{center}
\Ovalbox{Art 1}:
In this paper we formulate ${\mathcal N}=1$ supergravity as a constrained $BF$ theory with $OSp(4|1)$ gauge superalgebra. We derive the modified supergravity Lagrangian that, apart from the standard supergravity with negative cosmological constant, contains terms proportional to the (inverse of) Immirzi parameter. Although these terms do not change classical field equations, they might be relevant in quantum theory. We briefly discuss the perturbation theory around the supersymmetric topological vacuum.\\[9pt]
\Ovalbox{Art 2}:
We derive and analyze  Noether charges associated with the
diffeomorphism invariance for the constrained $SO(2,3)$ BF theory.
This result generalizes the Wald approach to the case of the first
order gravity with a negative cosmological constant, the Holst
modification and topological terms (Nieh-Yan, Euler, and
Pontryagin). We show that differentiability of the action is
automatically implemented by the the structure of the constrained BF
model. Finally, we calculate the AdS--Schwarzschild black hole
entropy from the Noether charge and we find that it
does not depend on the Immirzi parameter.\\[9pt]
\Ovalbox{Art 3}:
The framework of $SO(2,3)$ constrained BF theory applied to gravity
makes it possible to generalize formulas for gravitational diffeomorphic Noether charges (mass, angular momentum, and entropy). It extends Wald's approach to the case of first order gravity with a negative cosmological constant, the Holst modification and the topological terms (Nieh-Yan, Euler, and Pontryagin). Topological invariants play essential role contributing to the boundary terms in the regularization scheme for the asymptotically AdS spacetimes, so that the differentiability of the action is automatically secured. Intriguingly, it turns out that the black hole thermodynamics does not depend on the Immirzi parameter for the AdS--Schwarzschild, AdS--Kerr, and topological black holes, whereas a nontrivial modification appears for the AdS--Taub--NUT spacetime, where it impacts not only the entropy, but also the total mass.\\[9pt]
\Ovalbox{Art 4}:
We deform the anti-de Sitter algebra  by adding  additional generators $\mathcal{Z}_{ab}$,
forming in this way the negative cosmological constant counterpart
of the Maxwell algebra. We gauge this algebra and construct a
dynamical model with the help of a constrained the BF theory. It
turns out that the resulting theory is described by the
Einstein-Cartan action with Holst term, and the gauge fields
associated with the Maxwell generators $\mathcal{Z}_{ab}$ appear
only in topological terms that do not influence dynamical field
equations. We briefly comment on the extension of this construction,
which would lead to a nontrivial Maxwell fields dynamics.\\[9pt]
\Ovalbox{Art 5}:
In this paper we derive the Anti de Sitter counterpart of the
super-Maxwell algebra presented recently by Bonanos et.\ al. Then we
gauge this algebra and derive the corresponding supergravity theory,
which turns out to be described by the standard $\mathcal{N}=1$ supergravity
lagrangian, up to topological terms.\\[10pt]
\Ovalbox{Art 6}:
In this paper we discuss canonical analysis of $SO(4,1)$ constrained BF theory. The action of this theory contains  topological terms appended by a term that breaks the gauge symmetry down to the Lorentz subgroup of $SO(3,1)$. The equations of motion of this theory turn out to be the vacuum Einstein equations. By solving the $B$ field equations one finds that the action of this theory contains not only the standard Einstein-Cartan term, but also the Holst term proportional to the inverse of the Immirzi parameter, as well as a combination of topological invariants. We show that the structure of the constraints of a $SO(4,1)$ constrained BF theory is exactly that of gravity in Holst formulation.

%% file: Acknowledgments.tex
\textit{To my family, friends, mentors, and colleagues...\\[20pt]
Thank you all, who have crossed my path, shared a walk beside me, and will continue\\the trip toward the future.\\[15pt]
Without you, the journey I have just completed would not have brought me here, made me\\ who I am, and prepared me for the next part...}
\\[310pt]
\textit{I would like to thank my supervisor, Professor Jerzy Kowalski-Glikman, for his support and a fruitful collaboration during my studies}.\\[35pt]
\footnotesize
This PhD dissertation has been by supported by the grants of National Centre for Science Research (NN202112740 and 2011/01/N/ST2/00409), as well as the Max Born scholarship for PhD students supported by the city of Wroc\l aw and scholarships within the project: "Rozw\'{o}j potencja\l u i oferty edukacyjnej Uniwersytetu Wroc\l awskiego".

\newpage
\thispagestyle{empty}
~~
\newpage
\thispagestyle{empty}
~~\\[550pt]
\large
\begin{flushright}
    \textit{If we knew what it was we were doing, it would not be called research, would it?}
\end{flushright}
\begin{flushright}
   \normalsize \textsc{Albert Einstein} 
\end{flushright}
\newpage
\thispagestyle{empty}
\mbox{}